\documentclass[oneside,11pt]{article}

\usepackage[utf8]{inputenc}
\usepackage[T1]{fontenc}

\usepackage{amsthm}
\usepackage{amsmath}
\usepackage{amsfonts}
\usepackage{amssymb}

\usepackage{mathtools}

\usepackage{graphicx} 
\usepackage{float}
\usepackage{booktabs}
\usepackage{multirow}
\usepackage{rotating}
\usepackage{array}
\usepackage{tabularx}
\usepackage{widetable}

\usepackage{breakcites}

\usepackage{enumitem}

\usepackage[top=1.5cm,bottom=2cm,right=2.5cm,left=2.5cm]{geometry}
\usepackage{titling}

\usepackage{natbib}

\usepackage{ifplatform}

\ifwindows
\fi

\ifwindows
	\usepackage{bbm}
\else
	\iflinux
		\usepackage{bbm}
	\else
		\usepackage{bbm}
	\fi
\fi

\newcommand{\R}{\mathbb{R}}
\newcommand{\RR}{\mathbb{R}}

\newcommand{\PR}{\mathbb{P}}

\newcommand{\Sv}{\mathbb{S}}

\newcommand{\Hbb}{\mathbb{H}}

\newcommand{\lc}{\left[}
\newcommand{\rc}{\right]}
\newcommand{\lb}{\left\{}
\newcommand{\rb}{\right\}}

\newcommand{\ri}{\right.}

\newcommand{\prob}[1]{\PR\lb#1\rb}

\newcommand{\espe}[1]{\mathbb{E}\lc#1\rc}

\providecommand{\abs}[1]{\left|#1\right|}

\newcommand{\lrp}[1]{\left(#1\right)}
\newcommand{\lrc}[1]{\left[#1\right]}
\newcommand{\lrb}[1]{\left\{#1\right\}}
\newcommand{\E}[1]{\mathbb{E}\lc #1\rc}

\newcommand{\norm}[1]{\left|\left| #1\right|\right|}

\newcommand{\inprod}[2]{\left<#1,#2\right>}
\newcommand{\Xcal}{\mathcal{X}}

\newcommand{\ind}[1]{\mathbbm{1}_{\lrb{#1}}}

\usepackage[pdftex,final,bookmarksnumbered,bookmarksopen=false,breaklinks,colorlinks]{hyperref}
\hypersetup{
	final,
    bookmarksnumbered=true,
    bookmarksopen=true,
    bookmarksopenlevel=0,
    unicode=false,          
    pdftoolbar=true,        
    pdfmenubar=true,        
    pdffitwindow=false,     
    pdftitle={A goodness-of-fit test for the functional linear model with scalar response},    
	pdfdisplaydoctitle=true, 
	pdftoolbar=true,
	pdfmenubar=true,
	pdflang={English},
    pdfauthor={Eduardo Garcia-Portugues, Wenceslao Gonzalez-Manteiga, Manuel Febrero-Bande},     
    pdfsubject={arXiv paper},   
    pdfcreator={Eduardo Garcia-Portugues},   
    pdfproducer={Eduardo Garcia-Portugues}, 
    pdfkeywords={Bootstrap calibration}{Functional data}{Functional linear model}{Goodness-of-fit}, 
    pdfnewwindow=true,      
    breaklinks=true,
    hidelinks,       
    linkcolor=black,        
    citecolor=black,        
    filecolor=magenta,      
    urlcolor=cyan           
}

\newtheorem{lemma}{Lemma}

\allowdisplaybreaks

\begin{document}

\title{A goodness-of-fit test for the functional linear model\\ with scalar response}
\setlength{\droptitle}{-1cm}
\predate{}%
\postdate{}%
\author{Eduardo Garc\'ia-Portugu\'es$^{1,2}$, Wenceslao Gonz\'alez-Manteiga$^1$, and Manuel Febrero-Bande$^1$}

\date{}

\footnotetext[1]{
Department of Statistics and Operations Research, University of Santiago de Compostela (Spain).}
\footnotetext[2]{Corresponding author. e-mail: \href{mailto:eduardo.garcia@usc.es}{eduardo.garcia@usc.es}.}

\maketitle


\begin{abstract}
In this work, a goodness-of-fit test for the null hypothesis of a functional linear model with scalar response is proposed. The test is based on a generalization to the functional framework of a previous one, designed for the goodness-of-fit of regression models with multivariate covariates using random projections. The test statistic is easy to compute using geometrical and matrix arguments, and simple to calibrate in its distribution by a wild bootstrap on the residuals. The finite sample properties of the test are illustrated by a simulation study for several types of basis and under different alternatives. Finally, the test is applied to two datasets for checking the assumption of the functional linear model and a graphical tool is introduced. Supplementary materials are available online.
\end{abstract}

\begin{flushleft}
\end{flushleft}

\section{Introduction}

Functional data analysis has grown in popularity for the last years due to the increasingly data availability for continuous time processes. Typical examples of functional data include the temperature evolution, stock prices and path trajectories for objects in movement. New statistical methods have been developed to deal with the richer nature of functional data, being \cite{Ramsay2005}, \cite{Ferraty2006} and \cite{Ferraty2011} some of the main reference books in this area.\\

In many situations, the functional data is related to a scalar variable. For this cases, it is interesting to assess the relation  of the variables via a regression model, which can be used to predict the scalar response from the functional input. Analogue to the multivariate situation, the simplest functional regression model corresponds to the functional linear model with scalar response (see \cite{Ramsay2005} for a review). \\

An interesting methodology approach to deal with functional data is the use of random projections. The objective is to characterize the behaviour of a functional process, which has infinite dimension, via the behaviour of the one dimensional inner products of the functional process with suitable random functions. This method has interesting applications for the goodness-of-fit of the distribution of the process, as it can be seen in \cite{Cuesta-Albertos2007}. More recently, \cite{Patilea2012} provide a projection-based test for functional covariate effect in a functional regression model with scalar response. In their paper, the authors adapt the tests of \cite{Zheng1996} and \cite{Lavergne2008}, based on smoothing techniques, to the context of functional covariates. \\

In this work, a first goodness-of-fit test for the null hypothesis of the functional linear model, $H_0: m\in \lrb{\inprod{\cdot}{\beta}:\beta\in\Hbb}$, being $\Hbb$ the Hilbert space of square integrable functions, is proposed. The statistic test is of a  Cram\'er-von Mises type and is based on a generalization of a previous test of \cite{Escanciano2006}, designed for the case of a regression model with multivariate covariates. The test statistic is easy to compute using geometrical arguments and simple to calibrate in its distribution by a wild bootstrap on the residuals. Further, although the test is given for the functional linear model, it can be extended to other functional models with scalar response, as it is based on the residuals of the model. \\

This work is organized as follows. Some background on functional data, the functional linear model and the random projections paradigm are introduced in Section \ref{back}. The main part of this work is Section \ref{thetest}, where the theoretical arguments of the test, jointly with the bootstrap calibration procedure, are presented. The finite sample properties of the test are illustrated by a simulation study in Section \ref{simu}. Section \ref{appli} illustrates the application of the test to two datasets and introduces a graphical tool to evaluate the goodness-of-fit of the functional linear model with scalar response. Final comments and conclusions are given in Section \ref{final}. An appendix attached contains omitted proofs, tables and figures. 

\section{Background}
\label{back}

The main goal of this paper is to propose a goodness-of-fit test for the null hypothesis of the functional linear model with scalar response. Bearing in mind the different nature of the functional variables, some background on functional data, the functional linear model and the use of random projections is introduced. 

\subsection{Functional data}
\label{func}

One of the first and most important problems when we deal with functional data is to choose a suitable functional space to work. The most used functional spaces are the metric, the Banach and the Hilbert spaces. This is a sequence of functional spaces with increasing richer structure, where the tools available for the former space are included in the latter. Specifically, in a metric space we can measure distances between functions; in addition, in a Banach space we can also measure the functions and Cauchy sequences are convergent; and finally, in a Hilbert space we have inner product, which allows to consider functional basis.\\

While there are a lot of types of metrics and norm spaces, the $L^p$ spaces are one of the most used. The $L^p[0,1]$ space, $1\leq p<\infty$, is defined as the set of all functions $f:[0,1]\rightarrow\RR$ such that their norm $||f||_p=\big(\int_0^1 \abs{f(t)}^p\,dt\big)^{\frac{1}{p}}$ is finite. The choice of the interval $[0,1]$ is done only to fix the integration limits and other intervals can be considered without major changes. The most important $L^p$ space corresponds to $p=2$, because is the only which has an associated inner product $\inprod{\cdot}{\cdot}$ such that $||f||_p=\inprod{f}{f}^{\frac{1}{2}}$. For two functions $f,g\in L^2[0,1]$, their inner product is defined as
\begin{align*}
\inprod{f}{g}=\int_0^1 f(t)g(t)\,dt.
\end{align*}

In what follows we will consider as our working space the Hilbert space $\Hbb=L^2[0,1]$, bearing in mind that $[0,1]$ can be trivially replaced by another interval. The inner product allows for a basis representation of the elements of $\Hbb$ and, given a functional basis $\lrb{\Psi_j}_{j=1}^\infty$ of $\Hbb$, then any function $\Xcal$ in $\Hbb$ can be expressed by the linear combination $\Xcal=\sum_{j=1}^\infty x_j \Psi_j$, where $x_j=\inprod{\Xcal}{\Psi_j}$, $j\geq 1$. A basis is said to be orthogonal if $\inprod{\Psi_i}{\Psi_j}=0$, $i\neq j$ and orthonormal if, in addition, $\inprod{\Psi_j}{\Psi_j}=1$, $j\geq 1$. Typical examples of  basis of $\Hbb$ are the Fourier basis $\lrb{1,\sin\lrp{2\pi j x},\cos\lrp{2\pi j x}}_{j=1}^\infty$ and the B-splines basis (see \cite{Boor2001}).\\

For the development of the test statistic, we will also need to introduce a $p$-truncated basis $\lrb{\Psi_j}_{j=1}^p$, which corresponds to the first $p$ elements of the infinite basis $\lrb{\Psi_j}_{j=1}^\infty$. The representation of $\Xcal$ in this truncated basis is denoted by $\Xcal^{(p)}=\sum_{j=1}^p x_j \Psi_j$. The choice of the number of basis elements $p$ is crucial to have a reliable representation of the function $\Xcal$ by $\Xcal^{(p)}$. Although there exists several methods to select an appropriate $p$, we will refer to the GCV criteria (see \cite{Ramsay2005}, page 97) to select $p$ and represent adequately the function $\Xcal$ in $\lrb{\Psi_i}_{i=1}^p$. This criteria will be used in Section \ref{simu:simp} to select a suitable $p$ for the case of the simple hypothesis.\\

To deal with functional random projections we will need to define the functional analogue of the euclidean $p$-sphere $\Sv^p=\lrb{\mathbf{x}\in\RR^{p}:||\mathbf{x}||_{\RR^{p}}=1}$. In the functional case we have the \textit{functional sphere} of $\Hbb$, defined as $\Sv_\Hbb=\lrb{f\in\Hbb: ||f||_{\Hbb}=1}$, and the \textit{functional sphere of dimension $p$}, which is the set of functions of $\Hbb$ that, expressed in the $p$-truncated basis, have unit norm: $\Sv_\Hbb^p=\big\{f=\sum_{j=1}^p x_j\Psi_j\in\Hbb: ||f||_{\Hbb}=1\big\}$.\\

The relationship between $\Sv^p$ and $\Sv_\Hbb^p$ is particularly interesting to develop the test. Let be $\boldsymbol{\Psi}=\lrp{\inprod{\Psi_i}{\Psi_j}}_{ij}$ the matrix of inner products of the $p$-truncated basis, $\Sv_{\mathbf{\Psi}}^p=\lrb{\mathbf{x}\in\RR^p: \mathbf{x}^T\mathbf{\Psi}\mathbf{x}=1}$ the $p$-ellipsoid generated by this matrix and $\mathbf{R}^T\mathbf{R}$ the Cholesky decomposition of $\boldsymbol{\Psi}$ (a semi-positive matrix). First of all, we have the trivial isomorphism that maps elements of $\Sv_\Hbb^p$ to elements of $\Sv_{\mathbf{\Psi}}^p$ by means of the functional coefficients: $\phi: f=\sum_{j=1}^p x_j\Psi_j\in\Sv_\Hbb^p \mapsto \phi(f)=\mathbf{x}\in\Sv_{\mathbf{\Psi}}^p$. Recall that functions $\phi$ and $\phi^{-1}$ are well defined because $\norm{f}^2_{\Hbb}=\big\langle\sum_{j=1}^p x_j\Psi_j,\sum_{j=1}^p x_j\Psi_j\big\rangle=\mathbf{x}^T\mathbf{\Psi}\mathbf{x}$. We must consider also a linear transformation from $\Sv^p$ to $\Sv_{\mathbf{\Psi}}^p$, which is given by $\rho:\mathbf{x}\in\Sv^p\mapsto \rho(\mathbf{x})=\mathbf{R}^{-1}\mathbf{x}\in\Sv_{\mathbf{\Psi}}^p$ and whose Jacobian is $\abs{\mathbf{R}}^{-1}$, the determinant of the matrix $\mathbf{R}^{-1}$.\\
 
Using these two transformations, the integration of a functional operator $T$ with respect to a functional covariate $\gamma^{(p)}$ in $\Sv_\Hbb^p$ can be reduced to a real integration on the $p$-sphere:
\begin{align*}
\int_{\Sv_\Hbb^p} T\lrp{\gamma^{(p)}}d\gamma^{(p)}=\int_{\Sv_{\mathbf{\Psi}}^p} T\Bigg(\sum_{j=1}^p g_j \Psi_j\Bigg)\,d\mathbf{g}_{p}=\int_{\Sv^p} \abs{\mathbf{R}}^{-1}T\Bigg(\sum_{j=1}^p \lrp{\mathbf{R}^{-1}\mathbf{g}}_j \Psi_j\Bigg)\,d\mathbf{g}_{p}.
\end{align*}
In the case where the basis is orthonormal, $\mathbf{\Psi}$ and $\mathbf{R}$ are the identity matrix of order $p$. Then the coefficients of $\gamma^{(p)}\in\Sv^p_\Hbb$ in the basis $\lrb{\Psi_j}_{j=1}^p$ belong to $\Sv^p$ without any transformation.

\subsection{Functional linear model}
\label{funclm}

Suppose that $\Xcal$ is a functional random variable in $\Hbb$ and $Y$ is a real random variable. If both variables are centred, \textit{i.e.}, $\espe{\Xcal(t)}=0$ for a.e. $t\in[0,1]$ and $\espe{Y}=0$, the Functional Linear Model (FLM) with scalar response claims for the following relation:
\begin{align*}
Y=\inprod{\Xcal}{\beta}+\varepsilon=\int \Xcal(t)\beta(t)\,dt+\varepsilon,
\end{align*}
where the functional parameter $\beta$ belongs to $\Hbb$ and $\varepsilon$ is a random variable with zero mean, variance $\sigma^2$ and such that $\espe{\Xcal(t)\varepsilon}=0,\, \forall t$. The  prediction of $Y$ is done with the conditional expectation of $Y$ given $\Xcal$: 
\begin{align*}
m(\Xcal)=\espe{Y|\Xcal}=\inprod{\Xcal}{\beta}.
\end{align*}
Saying that $(\Xcal,Y)$ share the functional linear model is equivalent to saying that the regression function of $Y$ on $\Xcal$, $m$, belongs to the family $\mathcal{M}=\lrb{\inprod{\cdot}{\beta}:\beta\in\Hbb}$.\\

Given a sample $(\Xcal_1,Y_1),\ldots,(\Xcal_n,Y_n)$, the estimation of the functional parameter can be done by minimising the Residual Sum of Squares (RSS):
\begin{align*}
\hat\beta=\arg\min_{\beta\in\Hbb} \sum_{i=1}^n \lrp{Y_i-\inprod{\Xcal_i}{\beta}}^2.
\end{align*}
A possible method to search for the parameter $\beta$ that minimises the RSS is representing the functional data and the functional parameter in the truncated functional basis $\lrb{\Psi_j}_{j=1}^{p_\Xcal}$ and $\lrb{\theta_j}_{j=1}^{p_\beta}$, respectively: 
\begin{align*}
\Xcal_i^{(p_\Xcal)}=\sum_{j=1}^{p_\Xcal} c_{ij}\Psi_j,\, \beta^{(p_\beta)}=\sum_{j=1}^{p_\beta} b_j \theta_j,\, i=1,\ldots,n.
\end{align*}
Using the vector notation $\boldsymbol\Xcal=\big(\Xcal_i^{(p_\Xcal)}\big)_i$, $\mathbf{C}=(c_{ij})_{ij}$, $\boldsymbol \psi=(\Psi_j)_j$, $\mathbf{b}=(b_j)_j$ and $\boldsymbol\theta=(\theta_j)_j$, the previous representation can be expressed as $\boldsymbol\Xcal=\mathbf{C}\boldsymbol\psi$ and $\beta^{(p_\beta)}=\boldsymbol\theta^T \mathbf{b}$. The functional linear model results in
\begin{align}
Y=\inprod{\Xcal}{\beta}+\varepsilon\approx\mathbf{C}\mathbf{J} \mathbf{b}+\varepsilon=\mathbf{Zb}+\varepsilon,\label{modlin}
\end{align}
where $\mathbf{J}=\lrp{\inprod{\Psi_i}{\theta_j}}_{ij}$. Then, basis representation allows to express the FLM as a standard linear regression, where the estimated coefficients of $\beta$ in the basis $\lrb{\theta_j}_{j=1}^{p_\beta}$ are given by  $\hat{\mathbf{b}}=(\mathbf{Z}^T\mathbf{Z})^{-1}\mathbf{Z}^T\mathbf{Y}$.
Although different combinations of $\lrb{\Psi_j}_{j=1}^{p_\Xcal}$ and $\lrb{\theta_j}_{j=1}^{p_\beta}$ are possible, the usual choice is $\lrb{\Psi_j}_{j=1}^{p}=\lrb{\theta_j}_{j=1}^{p}$, being $\lrb{\Psi_j}_{j=1}^{p}$ an orthogonal basis because in that case the matrix $\mathbf{J}$ is diagonal.\\

There are several alternatives to represent the functional process and estimate the parameter $\beta$ in a truncated basis. For instance, a general review of the estimation based on the use of basis expansions such as Fourier series or B-splines can be found in the book by \cite{Ramsay2005}. The so called Functional Principal Component (FPC) regression estimation, proposed by \cite{Cardot1999}, provide an orthogonal data-driven basis that gives the most rapidly convergent representation of the functional dataset predictor in a $L^2$ sense (see \cite{Hall2007}). \cite{Preda2002} have proposed the Functional Partial Least Squares (FPLS) regression method that produces iteratively a sequence of orthogonal functions, as the FPC are, but with maximum predictive performance. To implement any of the methods shown before, it is required to fix the number of basis elements (or components) that are used in the estimation.\\

The optimal number of components, $p$, has to be fixed based on the information provided by the data. To do this, \cite{Hall2006} and \cite{Preda2002} use the predictive cross-validation criterion (PCV), \cite{Cardot2003} and \cite{Ferraty2011} consider the generalized cross-validation criterion (GCV) and \cite{Chiou2007} and \cite{Febrero-Bande2010} consider those methods based on the AIC, AICc and BIC information approaches.\\

Let denote by $\hat{Y}_i^{(p)}=\big\langle\Xcal_i^{(p)},\hat{\beta}^{(p)}\big\rangle$ and $\hat{Y}_{i,(-i)}^{(p)}=\big\langle\Xcal_i^{(p)},\hat{\beta}_{(-i)}^{(p)}\big\rangle$ the prediction of $Y_i$ using $p$ components with the whole sample and with the whole sample excluding the $i$-th element, respectively. The PCV is defined as:
\begin{align*}
\text{PCV}(p)=\arg\min_{p}\frac{1}{n}\sum_{i=1}^n \lrp{Y_i-\hat{Y}_{i,(-i)}^{(p)}}^2,
\end{align*} 
which is computationally expensive because it involves the estimation of the $\hat{\beta}_{(-i)}^{(p)}$ $n$ times. This is especially expensive in the case of data-driven basis (FPC, FPLS) because the basis has to be recalculate for every datum. As an alternative, GCV avoids recalculating the $\hat{\beta}^{(p)}$ for every datum by introducing a penalty term. The GCV is defined as
\begin{align}
\text{GCV}(p)=\arg\min_{p}\frac{\sum_{i=1}^n \lrp{Y_i-\hat{Y}_{i}^{(p)}}^2}{n\Big(1-\frac{df}{n}\Big)}, \label{gcv}
\end{align} 
where $df$ is the number of degrees of freedom consumed by the model, typically given by the trace of the matrix $\mathbf{Z}$. GCV is closely related with AIC, AICc and BIC although they come from different perspectives. 

\subsection{Random projections}

Random projections are becoming quite popular when dealing with high dimensional data, as a way to overcome the well known \textit{curse of the dimensionality}. The main idea behind is to reduce the dimension, and characterize the original distribution of the multidimensional data by the distribution of the randomly projected univariate data.\\

In the goodness-of-fit field, this is specially interesting, as the test procedures tend to become less efficient, less powerful, when the dimension of the model increases. \cite{Escanciano2006} used this technique to develop a goodness-of-fit test for multivariate regression models based on random projections. According to his simulation study, the test has an excellent power performance and has the best empirical power for most situations when comparing to their competitors.\\

In the functional framework, it is also possible to consider random projections. Usually, this is achieved by considering the inner product of the functional variable $\Xcal$ of $\Hbb$ and a suitable family of projectors, \textit{i.e.} random functions $\gamma$ in $\Hbb$. For example, using with this approach \cite{Cuesta-Albertos2007} developed some goodness-of-fit tests for parametric families of functional distributions, which includes goodness-of-fit tests for Gaussianity and for the Black--Scholes model.\\

A very interesting result on projections can be found in \cite{Patilea2012}. In their paper, the authors provide a characterization of the conditional expectation of a scalar variable $Y$ with respect to a functional variable $\Xcal$ given in terms of the conditional expectation of $Y$ with respect to the projected  $\Xcal$. The result is stated here in the following lemma. 

\begin{lemma}[\cite{Patilea2012}]
\label{l1}
Let $Y$ be a random variable and $\Xcal$ a functional random variable in the functional space $\Hbb$. The following statements are equivalent:
\begin{enumerate}[label=\roman{*}., ref=\roman{*}]

\item $\espe{Y|\Xcal=x}=0$, for almost every (a.e.) $x\in \Hbb$. \label{l1:1}

\item $\espe{Y|\inprod{\Xcal}{\gamma}=u}=0$, for a.e. $u\in \RR$ and $\forall\gamma\in \Sv_\Hbb$. \label{l1:2}

\item $\espe{Y|\inprod{\Xcal}{\gamma}=u}=0$, for a.e. $u\in \RR$ and $\forall\gamma\in \Sv^p_{\Hbb}$, $\forall p\geq 1$. \label{l1:3}
\end{enumerate}
\end{lemma}

\section{The test}
\label{thetest}

The presentation of the goodness-of-fit test that we propose in this paper is divided into three sections. The first and most important presents the theoretical fundamentals of the test, with starting point in Lemma \ref{p1}, which proof is detailed in the appendix. The second derives the effective implementation of the test statistic in practise considering some geometrical and matrix arguments. Finally, the bootstrap resampling for the calibration of the test statistic is presented in the last section.

\subsection{Theoretical arguments}

Let $Y$ be a real random variable and $\Xcal$ a functional random variable in the space $\Hbb$. Given a random sample $\lrb{\lrp{\Xcal_i,Y_i}}_{i=1}^n$, we are interested in checking if a functional linear model is suitable to explain the relation between the functional covariate and the scalar response, \textit{i.e.}, test for the composite hypothesis:
\begin{align*}
H_0: m\in \lrb{\inprod{\cdot}{\beta}:\beta\in\Hbb},
\end{align*}
versus a general alternative of the form $H_1: \prob{ m\notin \lrb{\inprod{\cdot}{\beta}:\beta\in\Hbb}}>0$. Further, the simple hypothesis, \textit{i.e.} checking for a specific functional linear model:
\begin{align*}
H_0: m\lrp{\Xcal}=\inprod{\Xcal}{\beta_0},\text{ for a fixed } \beta_0\in \Hbb,
\end{align*}
is also of interest as it includes the important case of no interaction between the functional covariate and the scalar response (considering $\beta_0(t)=0,\, \forall t$). In what follows we will focus on the procedure for the composite hypothesis, given that the simple is obtained just considering that the functional parameter is known and substituting $\hat\beta$ and $\hat\beta^{(p)}$ by $\beta_0$ and $\beta_0^{(p)}$, respectively. \\

The key point to test the null hypothesis $H_0$ is the following lemma, an adaptation of the Lemma \ref{l1} to our setting, which gives the characterization of $H_0$ in terms of the random projections of $\Xcal$. 
\begin{lemma}
\label{p1}
Let $\beta$ be an element of $\Hbb$. The following statements are equivalent:
\begin{enumerate}[label=\textit{\roman{*}}., ref=\textit{\roman{*}}]
\item $m(\Xcal)=\inprod{\Xcal}{\beta}$, $\forall\Xcal\in\Hbb$. \label{p1:1}

\item $\espe{Y-\inprod{\Xcal}{\beta}|\Xcal=x}=0$, for a.e. $x\in\Hbb$. \label{p1:2}

\item $\espe{Y-\inprod{\Xcal}{\beta}|\inprod{\Xcal}{\gamma}=u}=0$, for a.e. $u\in\R$ and $\forall\gamma\in\Sv_\Hbb$. \label{p1:3}

\item $\espe{Y-\inprod{\Xcal}{\beta}|\inprod{\Xcal}{\gamma}=u}=0$, for a.e. $u\in\R$ and $\forall\gamma\in\Sv_\Hbb^p$, $\forall p\geq 1$. \label{p1:3b}

\item $\espe{(Y-\inprod{\Xcal}{\beta})\mathbbm{1}_{\lrb{\inprod{\Xcal}{\gamma}\leq u}}}=0$, for a.e. $u\in\R$ and $\forall\gamma\in\Sv_\Hbb$. \label{p1:4}

\item $\espe{(Y-\inprod{\Xcal}{\beta})\mathbbm{1}_{\lrb{\inprod{\Xcal}{\gamma}\leq u}}}=0$, for a.e. $u\in\R$ and $\forall\gamma\in\Sv_\Hbb^p$, $\forall p\geq 1$. \label{p1:4b}
\end{enumerate}
\end{lemma}

Then $H_0$ is characterized by the null value of the moment $\espe{(Y-\inprod{\Xcal}{\beta})\mathbbm{1}_{\lrb{\inprod{\Xcal}{\gamma}\leq u}}}$, for a.e. $u\in\RR$ and $\forall\gamma\in\Sv_\Hbb$ (or $\forall\gamma\in\Sv_\Hbb^p$, $\forall p\geq 1$) and a possible way to measure the deviation of the data from $H_0$ is by the empirical process arising from the estimation of this moment: 
\begin{align}
R_n(u,\gamma)=n^{-\frac{1}{2}}\sum_{i=1}^n \lrp{Y_i-\inprod{\Xcal_i}{\hat\beta}}\mathbbm{1}_{\lrb{\inprod{\Xcal_i}{\gamma}\leq u}},\label{rmpp}
\end{align}
that will be denoted as the \textit{Residual Marked empirical Process based on Projections} (RMPP). The marks of (\ref{rmpp}) are given by the residuals $\big\{Y_i-\big<\Xcal_i,\hat\beta\big>\big\}_{i=1}^n$ and the jumps by the projected functional regressor in the direction $\gamma$, $\lrb{\inprod{\Xcal_i}{\gamma}}_{i=1}^n$. 
The estimation of $\beta$ can be done by different methods as described in Section \ref{back}. Note that the RMPP only depends on the residuals of the model considered (in this case the residuals of the FLM) and therefore it can be easily extended to other regression models (see Section \ref{final} for discussion).\\

To measure the distance of the empirical process (\ref{rmpp}) from zero, two possibilities are the classical Cram\'er--von Mises and Kolmogorov--Smirnov norms, adapted to the \textit{projected} space $\Pi=\RR\times\Sv_{\Hbb}$:
\begin{align}
\text{PCvM}_n&=\int_\Pi R_n(u,\gamma)^2 \,F_{n,\gamma}(du)\,\omega(d\gamma),\label{cvm}\\
\text{PKS}_n&=\sup_{(u,\gamma)\in\Pi}\abs{R_n(u,\gamma)}\label{ks},
\end{align}
where $F_{n,\gamma}$ is the empirical cumulative distribution function (ecdf) of the projected functional data in the direction $\gamma$ (\textit{i.e.} the ecdf of the data $\lrb{\inprod{\Xcal_i}{\gamma}}_{i=1}^n$) and $\omega$ represents a measure on $\Sv_\Hbb$. Unfortunately, the infinite dimension of the space $\Sv_\Hbb$ makes infeasible to compute the functionals (\ref{cvm}) and (\ref{ks}) and some kind of discretization is needed. A solution to this problem is to consider the properties of the Hilbert space $\Hbb$ and use a basis representation.\\

Up to this end, let us introduce some notation. Let $\lrb{\Psi_j}_{j=1}^\infty$ be a basis of $\Hbb$ and consider the $p$-truncated basis $\lrb{\Psi_j}_{j=1}^p$, with matrix of inner products $\mathbf{\Psi}$. Denote by $\Xcal_i^{(p)}$ and $\gamma^{(p)}$ the representation of the functions $\Xcal_i$ and $\gamma$ in the $p$-truncated basis, with vectors of coefficients $\mathbf{x}_{i,p}$ and $\mathbf{g}_{p}$, respectively, and for $i=1,\ldots,n$. Using this, as $\lrb{\Psi_j}_{j=1}^\infty$ is any basis, we have that
\begin{align*}
\inprod{\Xcal_i^{(p)}}{\gamma^{(p)}}=\mathbf{x}_{i,p}^T\,\mathbf{\Psi}\,\mathbf{g}_{p}.
\end{align*}
By analogy with the previously defined $F_{n,\gamma}$, we will denote $F_{n,\gamma^{(p)}}$ to the ecdf of the projected functional data expressed in the $p$-truncated basis, both for the projector $\gamma$ and for the functional data. Then, the RMPP can be expressed in terms of a $p$-truncated basis, yielding
\begin{align*}
R_{n,p}\lrp{u,\gamma^{(p)}}%
&=n^{-\frac{1}{2}}\sum_{i=1}^n \lrp{Y_i-\mathbf{x}_{i,p}^T\,\mathbf{\Psi}\,\mathbf{b}_{p}}\mathbbm{1}_{\lrb{\mathbf{x}_{i,p}^T\,\mathbf{\Psi}\,\mathbf{g}_{p}\leq u}}=R_{n,p}\lrp{u,\mathbf{g}_p},
\end{align*}
where $\mathbf{b}_{p}$ represents the coefficients of $\hat\beta$ in the $p$-truncated basis $\lrb{\Psi_j}_{j=1}^p$. \\

Bearing in mind this, our test statistic propose is a modified version of (\ref{cvm}) that results from expressing all the functions in a $p$-truncated basis of $\Hbb$:
\begin{align}
\text{PCvM}_{n,p}&=\int_{\Sv_\Hbb^p\times\RR} R_{n,p}\lrp{u,\gamma^{(p)}}^2\,F_{n,\gamma^{(p)}}(du)\,\omega\big(d\gamma^{(p)}\big).\label{pcvmp1}
\end{align}
We have decided to choose the Cram\'er-von Mises statistic because, as we will see, presents important computational advantages and can be adapted to the given framework of \cite{Escanciano2006} for the finite dimensional case. The most important advantage is that we can derive an explicit expression where there is no need to compute the RMPP for different projections, property that does not hold for the Kolmogorov--Smirnov statistic.\\

Using that the integration in the $p$-sphere of $\Hbb$ can be expressed as the integration in the $p$-sphere of $\RR^{p}$ via the transformations defined in Section \ref{func}, we have:
\begin{align}
\text{PCvM}_{n,p}&=\int_{\Sv^p_{\boldsymbol\Psi}\times\RR} R_{n,p}(u,\mathbf{g}_{p})^2 \,F_{n,\mathbf{g}_{p}}(du)\,\omega(d\mathbf{g}_{p})\nonumber\\
&=\int_{\Sv^p\times\RR} \abs{\mathbf{R}}^{-1}R_{n,p}(u,\mathbf{R}^{-1}\mathbf{g}_{p})^2 \,F_{n,\mathbf{R}^{-1}\mathbf{g}_{p}}(du)\,\omega(d\mathbf{g}_{p})\nonumber\\
&=\int_{\Sv^p\times\RR} \abs{\mathbf{R}}^{-1}\lrp{n^{-\frac{1}{2}}\sum_{i=1}^n \lrp{Y_i-\mathbf{x}_{i,p}^T\,\mathbf{\Psi}\,\mathbf{b}_{p}}\mathbbm{1}_{\lrb{\mathbf{x}_{i,p}^T\,\mathbf{R}^T\,\mathbf{g}_{p}\leq u}}}^2 \,F_{n,\mathbf{R}^{-1}\mathbf{g}_{p}}(du)\,\omega(d\mathbf{g}_{p}),\label{pcvmp2}
\end{align}
where $\omega$ now represents a measure in the $p$-sphere $\Sv^p$ that, for simplicity purposes, will be considered as the uniform distribution on $\Sv^p$. \\

Essentially, what we have done is to treat the functional process as a $p$-multivariate process, expressing the functions in a basis of $p$ elements. The methods to choose the number of elements $p$ and to estimate the parameter $\beta$ both for the simple and for the composite hypothesis are the ones introduced in Section \ref{back}. These methods will be illustrated in Section \ref{simu}.

\subsection{Implementation}
\label{imple}

Following the steps of \cite{Escanciano2006} it is possible to derive a simpler expression for (\ref{pcvmp2}). Using the definition of the RMPP in a $p$-truncated basis, the fact that $F_{n,\mathbf{R}^{-1}\mathbf{g}_{p}}$ is the ecdf of $\big\{\mathbf{x}_{i,p}^T\mathbf{\Psi}\mathbf{R}^{-1}\mathbf{g}_{p} \big\}_{i=1}^n=\big\{\mathbf{x}_{i,p}^T\mathbf{R}^T\mathbf{g}_{p} \big\}_{i=1}^n$ and some simple algebra, we have:
\begin{align*}
\text{PCvM}_{n,p}&=\int_{\Sv^p\times\RR} \abs{\mathbf{R}}^{-1} R_{n,p}(u,\mathbf{R}^{-1}\mathbf{g}_{p})^2 \,F_{n,\mathbf{R}^{-1}\mathbf{g}_p}(du)\,d\mathbf{g}_{p}\\
&=n^{-1}\sum_{i=1}^n\sum_{j=1}^n \hat\varepsilon_i \hat\varepsilon_j\int_{\Sv^p\times\RR} \abs{\mathbf{R}}^{-1}\ind{\mathbf{x}_{i,p}^T\mathbf{R}^T\mathbf{g}_{p}\leq u}\ind{\mathbf{x}_{j,p}^T\mathbf{R}^T\mathbf{g}_{p}\leq u} \,F_{n,\mathbf{R}^{-1}\mathbf{g}_{p}}(du)\,d\mathbf{g}_{p}\\
&=n^{-2}\sum_{i=1}^n\sum_{j=1}^n\sum_{r=1}^n \hat\varepsilon_i \hat\varepsilon_j A_{ijr},
\end{align*}
with $\hat\varepsilon_i=Y_i-\big<\Xcal^{(p)}_i,\hat\beta^{(p)}\big>$. The terms $A_{ijr}$ represent the integrals
\begin{align*}
A_{ijr}=&\int_{\Sv^p} \abs{\mathbf{R}}^{-1}\ind{\mathbf{x}_{i,p}^T\mathbf{R}^T\mathbf{g}_{p}\leq \mathbf{x}_{r,p}^T\mathbf{R}^T\mathbf{g}_{p}}\ind{\mathbf{x}_{j,p}^T\mathbf{R}^T\mathbf{g}_{p}\leq \mathbf{x}_{r,p}^T\mathbf{R}^T\mathbf{g}_{p}}\,d\mathbf{g}_{p}\\
=&\int_{\Sv^p} \abs{\mathbf{R}}^{-1}\ind{(\mathbf{R}\mathbf{x}_{i,p}-\mathbf{R}\mathbf{x}_{r,p})^T\mathbf{g}_{p}\leq 0 ,\,(\mathbf{R}\mathbf{x}_{j,p}-\mathbf{R}\mathbf{x}_{r,p})^T\mathbf{g}_{p}\leq 0}\,d\mathbf{g}_{p}\\
=&\abs{\mathbf{R}}^{-1}\int_{S_{ijr}} d\mathbf{g}_{p},%
\end{align*}
where $S_{ijr}=\big\{\boldsymbol\xi\in\Sv^p: \frac{\pi}{2}\leq \measuredangle\big(\mathbf{x}'_{i,p}-\mathbf{x}'_{r,p},\boldsymbol\xi\big)\leq\frac{3\pi}{2},\,\frac{\pi}{2}\leq \measuredangle\big(\mathbf{x}'_{j,p}-\mathbf{x}'_{r,p},\boldsymbol\xi\big)\leq\frac{3\pi}{2}\big\}$ and $\measuredangle\lrp{\mathbf{a},\mathbf{b}}$ represents the angle between vectors $\mathbf{a}$ and $\mathbf{b}$. To simplify notation, we denote $\mathbf{x}'_{k,p}=\mathbf{R}\mathbf{x}_{k,p}$ ($\mathbf{x}'_{k,p}=\mathbf{x}_{k,p}$ if the basis is orthonormal) for $k=1,\ldots,n$. Depending on $\mathbf{x}'_{i,p},\,\mathbf{x}'_{j,p},\,\mathbf{x}'_{r,p}$, the region $S_{ijr}$ can be the whole sphere $\Sv^p$ ($\mathbf{x}'_{i,p}=\mathbf{x}'_{j,p}=\mathbf{x}'_{r,p}$), a hemisphere of $\Sv^p$ ($\mathbf{x}'_{i,p}\neq\mathbf{x}'_{j,p}$, $\mathbf{x}'_{i,p}=\mathbf{x}'_{r,p}$ or $\mathbf{x}'_{j,p}=\mathbf{x}'_{r,p}$) or a spherical wedge (see Figure \ref{fig0} in appendix) of width angle given by
\begin{align}
\abs{\pi-\arccos\lrp{\frac{(\mathbf{x}'_{i,p}-\mathbf{x}'_{r,p})^T(\mathbf{x}'_{j,p}-\mathbf{x}'_{r,p})}{||\mathbf{x}'_{i,p}-\mathbf{x}'_{r,p}||\cdot||\mathbf{x}'_{j,p}-\mathbf{x}'_{r,p}||}}}. \label{angle}
\end{align}

Thus $A_{ijr}$ is the product of the surface area of a spherical wedge of angle $A^{(0)}_{ijr}$ times $\abs{\mathbf{R}}^{-1}$, and is given by
\begin{align*}
A_{ijr}=A_{ijr}^{(0)}\frac{\pi^{{p/2}-1}}{\Gamma\lrp{\frac{p}{2}}}\abs{\mathbf{R}}^{-1},
\quad A_{ijr}^{(0)}=\lb\begin{array}{ll}
2\pi, & \mathbf{x}'_{i,p}=\mathbf{x}'_{j,p}=\mathbf{x}'_{r,p},\\
\pi, & \mathbf{x}'_{i,p}\neq\mathbf{x}'_{j,p}, \mathbf{x}'_{i,p}=\mathbf{x}'_{r,p}\text{ or }\mathbf{x}'_{j,p}=\mathbf{x}'_{r,p},\\
(\ref{angle}), & \text{else.}
\end{array}\ri
\end{align*}

We also have a symmetric property, $A_{ijr}=A_{jir}$, which simplifies the evaluation of the test statistic from $\mathcal{O}(n^3)$ to $\mathcal{O}\lrp{(n^3+n^2)/2}$ computations. The memory requirement is expensive, because we need to store the $(n^3+n^2)/2$ elements of the three dimensional array $\mathbf{A}$, which is symmetric in its two first indexes. However, this requirement can be stretched if we consider the following expression for the statistic:
\begin{align}
\text{PCvM}_{n,p}=n^{-2}\hat{\boldsymbol\varepsilon}^T \mathbf{A}_{\bullet} \hat{\boldsymbol\varepsilon},\label{efficpcvm}
\end{align}
where $\mathbf{A}_{\bullet}=\lrp{\sum_{r=1}^n A_{ijr}}_{ij}$ is a $n\times n$ matrix and $\hat{\boldsymbol\varepsilon}$ is the vector of the residuals. By the definition of $A_{ijr}^{(0)}$ and its symmetry in the first two entries, the matrix $\mathbf{A}_{\bullet}$ is symmetric and its diagonal terms are given by $(n+1)\pi$. Although the order of computations remains similar, $\mathcal{O}\lrp{(n^3-n^2)/2}$, the memory required for storing the matrix $\mathbf{A}_{\bullet}$ is substantially lower and drops to $(n^2-n+2)/2$ elements. This fact improves drastically the time of computation of the statistic and allows to\nopagebreak[4] apply the test to larger datasets.\\

Again, let us remark that the expression derived for the $\text{PCvM}_{n,p}$ statistic remains valid for any functional regression model with scalar response and not just for the FLM, as the expression is based on the residuals of the model.

\subsection{Bootstrap resampling}

To calibrate the distribution of the statistic $\text{PCvM}_{n,p}$ under the null hypothesis, a wild bootstrap on the residuals is applied. This bootstrap procedure is consistent in the finite dimensional case, as it was shown in \cite{Stute1998}, and is adequate to situations with potential heterocedasticity, quite common in functional data. The resampling process for the case of the composite hypothesis, given an initial estimation $\hat\beta^{(p)}$ of the functional parameter, is the following:

\begin{enumerate}[label=\textit{\roman{*}}., ref=\textit{\roman{*}}]

\item Construct the estimated residuals: $\hat\varepsilon_i=Y_i-\big\langle\Xcal_i^{(p)},\hat\beta^{(p)}\big\rangle,\,i=1,\ldots,n$. \label{boot1}
\item Draw independent random variables $V_1^*,\ldots,V_n^*$ satisfying $\mathbb{E}^*\lrc{V_i^{*}}=0\text{ and }\mathbb{E}^*\lrc{V_i^{*2}}=1$. For example, if $V^*$ is a discrete random variable with distribution weights
$\mathbb{P}\big\{V^*=\frac{1-\sqrt{5}}{2}\big\}=\frac{5+\sqrt{5}}{10}$ $\mathbb{P}\big\{V^*=\frac{1+\sqrt{5}}{2}\big\}=\frac{5-\sqrt{5}}{10}$, we have the \textit{golden section bootstrap}.\label{boot2}

\item Construct the bootstrap residuals $\varepsilon^*_i=V_i^*\hat\varepsilon_i,\,i=1,\ldots,n$.\label{boot3}

\item Set $Y_i^*=\big\langle\Xcal_i^{(p)},\hat\beta^{(p)}\big\rangle+\varepsilon_i^*,\,i=1,\ldots,n$ and estimate $\beta^{*,(p)}$ for the sample $\lrb{\lrp{\Xcal_i,Y_i^*}}_{i=1}^n$.\label{boot4}

\item Obtain the estimated bootstrap residuals $\hat\varepsilon_i^*=Y_i^*-\big\langle\Xcal_i^{(p)},\hat\beta^{*,(p)}\big\rangle,\,i=1,\ldots,n$.\label{boot5}

\end{enumerate}

Then, the procedure to calibrate the test is the following. In step \ref{boot1} we compute the test statistic with the residuals under $H_0$ using the implementation (\ref{efficpcvm}) of the previous section. Then repeat steps \ref{boot2}--\ref{boot5} for $b=1,\ldots,B$, computing each time the bootstrap statistic $\text{PCvM}^{*,b}_{n,p}=n^{-2}\hat{\boldsymbol\varepsilon}^{*,b,T} \mathbf{A}_{\bullet} \hat{\boldsymbol\varepsilon}^{*,b}$ and estimate the $p$-value of the test by Monte Carlo: $\#\lrb{\text{PCvM}_{n,p}\leq\text{PCvM}^{*,b}_{n,p}}/B$. For computational efficiency, it is important to note that we do not have to compute again the matrix $\mathbf{A}_{\bullet}$ in the bootstrap replicates. \\

A very interesting fact of the FLM is that step \ref{boot5} can be easily performed using the properties of the estimation of $\hat\beta^{(p)}$. From (\ref{modlin}) it is clear that the vector of coefficients of $\hat\beta^{(p)}$ is estimated throughout $\hat{\mathbf{b}}=\lrp{\mathbf{Z}^T\mathbf{Z}}^{-1}\mathbf{Z}^T\mathbf{Y}$. Then, the estimated bootstrap residuals, represented by the vector $\hat{\boldsymbol\varepsilon}^*$, can be obtained as $\hat{\boldsymbol\varepsilon}^*=\big(\mathbf{I}_{p}-\mathbf{Z}\lrp{\mathbf{Z}^T\mathbf{Z}}^{-1}\mathbf{Z}^T\big)\mathbf{Y}^*$, where $\mathbf{Y}^*$ is the vector of bootstrap responses given in step \ref{boot4} and $\mathbf{I}_{p}$ is the identity matrix of order $p$. The projection matrix $\big(\mathbf{I}_{p}-\mathbf{Z}\lrp{\mathbf{Z}^T\mathbf{Z}}^{-1}\mathbf{Z}^T\big)$ remains the same for all the bootstrap replicates, so it can be stored without the need of computing it again. Obtaining the residuals in this way implies a significative computational saving.\\

The bootstrap resampling in the case of the simple hypothesis is easier: just replace $\hat\beta^{(p)}$ by $\beta_0^{(p)}$ and omit steps \ref{boot4} and \ref{boot5}, considering $\hat\varepsilon_i^*=\varepsilon_i^*,\,i=1,\ldots,n$.

\section{Simulation study}
\label{simu}

To illustrate the finite sample properties of the proposed test, a simulation study was carried out for the simple and the composite hypotheses. The functional process considered for the functional covariate $\Xcal$ is an Ornstein--Uhlenbeck process in $[0,1]$, which corresponds to a Brownian motion with functional mean $\mu$ and covariance function given by $\text{Cov}(\Xcal(s),\Xcal(t))=\frac{\sigma^2}{2\theta}e^{-\theta(s+t)}\lrp{e^{2\theta\min(s,t)}-1}$.
We have considered $\theta=\frac{1}{3}$, $\sigma=1$ and the functional mean $\mu(t)=0$, $\forall t\in[0,1]$. See Figure \ref{fig1} in appendix for further details. \\

All the functional data in this simulation study is represented in $201$ equidistant points in the interval $[0,1]$. The number of bootstrap replicates considered is $B=1000$ and the number of Monte Carlo replicates for determining the empirical sizes and powers, $M=1000$. The sample size, except otherwise stated, is $n=100$. Lastly, in order to properly compare the effect of the kind of basis, the number of elements and the sample sizes, the initial seed for the random generation of the functional underlying process is the same for each model.\\

Several lengthy tables have been reduced in this section for space saving. The reader is referred to the appendix to see the whole tables as well as other explanatory figures. 

\subsection{Testing for simple hypothesis}
\label{simu:simp}

The simulation study for the simple hypothesis is centred on the case $H_0: m(\Xcal)=\inprod{\Xcal}{\beta_0}$, where $\beta_0(t)=0,\,t\in[0,1]$. This is equivalent to test that the functional covariate $\Xcal$ has no effect on the scalar response, \textit{i.e.}, test the null hypothesis $H_0: m(\Xcal)=0$. Although there is an extensive collection of goodness-of-fit tests for finite dimensional covariates (see \cite{Gonzalez-Manteiga2011}), the literature for the case of functional covariates is more limited. Therefore, we will focus on the competing procedures of \cite{Delsol2011} and \cite{Gonzalez-Manteiga:bootflm} to compare the different tests in terms of level and power. Let us describe briefly these two test statistics.\\

\cite{Delsol2011} propose a test statistic for $H_0: m(\Xcal)=m_0(\Xcal)$, deriving its asymptotic law and giving a bootstrap procedure based on the residuals. The statistic, inspired in the propose of \cite{Hardle1993}, is
\begin{align*}
T_n=\int\lrp{\sum_{i=1}^n (Y_i-m_0(\Xcal_i))K\lrp{\frac{d(\Xcal,\Xcal_i)}{h}}}^2\omega(\Xcal)\,dP_\Xcal(\Xcal),
\end{align*}
where $K$ is a kernel function, $d$ is a semimetric and $h$ is the bandwidth parameter. $P_\Xcal$ represents the probability distribution of the functional process and  $\omega$ is a suitable weight function. The test used in our implementation results from considering no functional effect, \textit{i.e.} $H_0: m_0(\Xcal)=0$, and from approximating the integral with respect to $dP_\Xcal$ by the empirical mean of the sample. We have also considered the kernel $K(t)=2\phi(\abs{t}),\,t\in\RR$, being $\phi$ the density of a $\mathcal{N}(0,1)$, the $L^2$ distance in $\Hbb$ for $d$ and the uniform weight function. The bandwidth parameter is given by the PCV criterion and bootstrap resampling was done using golden wild bootstrap. \\

The other competing test is the one proposed by \cite{Gonzalez-Manteiga:bootflm} and is based on the idea of extending the covariance to functional-scalar data:
\begin{align*}
D_n=\norm{\frac{1}{n}\sum_{i=1}^n \lrp{\Xcal_i-\bar\Xcal}\lrp{Y_i-\bar Y}}_\Hbb,
\end{align*}
where $\bar\Xcal$ is the functional mean of $\lrb{\Xcal_i}_{i=1}^n$ and is $\bar Y$ the usual scalar mean of $\lrb{Y_i}_{i=1}^n$. The authors extend the ideas of the classical $F$-test to the functional framework,  resulting a statistic to test the null hypothesis of no interaction \textit{inside} the functional linear model. The test is consistent and the authors derived the asymptotic distribution of the process $\frac{1}{n}\sum_{i=1}^n \lrp{\Xcal_i-\bar\Xcal}\lrp{Y_i-\bar Y}$, resulting in a Brownian motion with mean $\E{(\Xcal-\mu_\Xcal)(Y-\mu_Y)}$ and a particular covariance structure. This test can be viewed as a possible benchmark in our simulation study and, recalling its similarity with the classical $F$-test, will be denoted as the \textit{functional $F$-test}. The bootstrap resampling was also performed using golden wild bootstrap.\\

Three different blocks of deviations from the null are considered. The first two blocks represent a deviation inside the linear model, \textit{i.e.}, considering different functions $\beta_{j,k},\,j=1,2,\,k=1,2,3,$ instead of $\beta_0$. The linear functions are $\beta_{1,k}(t)=\gamma_k(t-0.5)$, with coefficients $\gamma_1=0.25$, $\gamma_2=0.65$ and $\gamma_3=1.00$ for $H_{1,k}$ and $\beta_{2,k}(t)=\eta_k\sin(2\pi t^3)^3$, with $\eta_1=0.10$, $\eta_2=0.20$ and $\eta_3=0.50$ for $H_{2,k}$. The second block of deviations from the null hypothesis consists on adding a \textit{second order} term $\inprod{\Xcal}{\Xcal}$ to the regression function, thus the model is no longer linear. Different weights for the second term are represented in the alternatives
$H_{3,k}: Y=\inprod{\Xcal}{\beta_{0}}+\delta_k\inprod{\Xcal}{\Xcal}+\varepsilon$, where $\delta_1=0.005$, $\delta_2=0.010$ and $\delta_3=0.015$. The relation between the variance of the response with respect to the variance of the error can be measured by the \textit{signal-to-noise ratio}: $\text{snr}=\sigma^2/\lrp{\sigma^2+\E{m(\Xcal)^2}}$. For block 1 the snr's of the alternatives are $0.956$, $0.765$ and $0.579$, respectively for $H_{1,k}$, $k=1,2,3$. For block 2, the snr's are $0.981$, $0.850$ and $0.671$. For block 3, we have $0.985$, $0.914$ and $0.728$.\\

In the case of the simple hypothesis there is no estimation of the parameter $\beta_0$, as it is known. However, it is necessary to express the functional process $p$ and the function $\beta_0$ in a suitable basis in order to compute the test statistic. Up to this end, we consider a B-splines basis and we choose automatically its number of elements by the GCV criteria commented in Section \ref{func}.\\

The results of the study for the simple hypothesis are collected in Table \ref{tab1}, which shows the empirical sizes and powers of the functional $F$-test, the test of Delsol et al. and the PCvM test for simple hypothesis, for the models previously commented. All of the tests seem to calibrate the significance level $\alpha=0.05$. With respect to the power, the functional $F$-test has in average a superior behaviour in the alternatives $H_{2,k}$, which represents deviations from the null \textit{inside} a linear model. The test of Delsol et al. performs also well with the cross-validatory bandwidth, being the most competitive for the block $H_{1,k}$. The PCvM test performs worse than the functional $F$-test for alternatives $H_{1,k}$ and $H_{2,k}$ and similarly to the test of Delsol et al. Nevertheless, for alternatives that are not inside the linear model, the PCvM test results the most powerful. Similar results are obtained with a noise given by a recentred exponential distribution with parameter $\lambda=10$.

\begin{table}[!h]
\centering
\small
\begin{tabular}{c|ccc||ccc}\toprule\toprule
Models & $F$-test & PCvM & Delsol et al. & $F$-test & PCvM & Delsol et al. \\\midrule
$H_0$ & $0.060$ & $0.041$ & $0.065$ & $0.043$ & $0.051$ & $0.066$ \\\midrule
$H_{1,1}$ & $0.060$ & $0.069$ & $0.098$ & $0.056$ & $0.052$ & $0.072$ \\
$H_{1,2}$ & $0.163$ & $0.078$ & $0.309$ & $0.180$ & $0.085$ & $0.285$ \\
$H_{1,3}$ & $0.401$ & $0.138$ & $0.772$ & $0.442$ & $0.166$ & $0.719$ \\\midrule
$H_{2,1}$ & $0.248$ & $0.053$ & $0.080$ & $0.265$ & $0.071$ & $0.089$ \\
$H_{2,2}$ & $0.951$ & $0.336$ & $0.403$ & $0.932$ & $0.343$ & $0.420$ \\
$H_{2,3}$ & $1.000$ & $0.904$ & $0.877$ & $0.999$ & $0.901$ & $0.848$\\\midrule
$H_{3,1}$ & $0.034$ & $0.173$ & $0.165$ & $0.052$ & $0.125$ & $0.128$\\
$H_{3,2}$ & $0.038$ & $0.691$ & $0.554$ & $0.034$ & $0.721$ & $0.558$ \\
$H_{3,3}$ & $0.019$ & $0.998$ & $0.932$  & $0.012$ & $1.000$ & $0.967$ \\\bottomrule\bottomrule
\end{tabular}
\caption{\small Empirical power of the competing tests for the simple hypothesis $H_0:m(\Xcal)=\inprod{\Xcal}{\beta_0}$, $\beta_0(t)=0,\, \forall t$ and significance level $\alpha=0.05$. Noise follows a $\mathcal{N}\big(0,0.10^2\big)$ and a recentred $\text{Exp}(10)$.\label{tab1}}
\end{table}

\subsection{Testing for composite hypothesis}
\label{simu:comp}

To see the performance of the test under the composite hypothesis $H_0: m\in \lrb{\inprod{\cdot}{\beta}:\beta\in\Hbb}$ we have considered three different null models of the form
\begin{align}
H_{j,0}:\quad Y=\inprod{\Xcal}{\beta_{j}}+\varepsilon,\label{j0}
\end{align}
with $j=1,2,3$ being the index of the three different models. The functional coefficients of the three FLM are $\beta_{1}(t)=\sin(2\pi t)-\cos(2\pi t)$, $\beta_{2}(t)=t-\lrp{t-0.75}^2$ and $\beta_{3}(t)=t+\cos(2\pi t)$, $t\in[0,1]$. The second functional coefficient is chosen to be perfectly described by B-splines, whereas this is not the case for $\beta_1$ and $\beta_3$.\\

In order to check the power performance of the test, a set of possible deviations from the linear regression model is considered. Again, a second order term $\inprod{\Xcal}{\Xcal}$ is introduced to transform the model into a non-linear one. Three different weights for this term are considered, representing the alternatives $H_{j,k}$:
\begin{align}
 H_{j,k}:\quad Y=\inprod{\Xcal}{\beta_{j}}+\delta_k\inprod{\Xcal}{\Xcal}+\varepsilon.\label{jk}
\end{align}
The index for the model is denoted by $j=1,2,3$ and $k=1,2,3$  is the index that measures the degree of the deviation from the null hypothesis. The weights of the quadratic term are $\delta_1=0.01$, $\delta_2=0.05$ and $\delta_3=0.10$. The snr's for model 1 are $0.177$, $0.176$, $0.166$ and $0.140$, respectively for $H_{1,k}$, $k=0,1,2,3$. For model 2, $0.050$, $0.050$, $0.050$ and $0.047$. For model 3, we have $0.029$, $0.029$, $0.029$ and $0.028$.\\

Three estimation methods for the functional parameter $\beta$ will be considered. All of them are designed in order to provide automatic selectors of the number of elements considered in the basis estimation of $\beta$. So, the first automatic method considered is the estimation of $\beta$ as a linear combination of a B-splines basis of $p$ elements, where $p$ is chosen by the GCV criteria (\ref{gcv}). Secondly, FPC estimation relies on the BIC criteria to choose the optimal number of elements in the FPC basis derived from the process to estimate $\beta$. Finally, the FPLS method also uses PCV to select the adequate number of elements in the FPLS basis derived from the joint sample \nolinebreak[4] $\lrb{\lrp{\Xcal_i,Y_i}}_{i=1}^n$.\\

Table \ref{tab4new} shows the rejection frequencies of the null hypothesis for the test computed from observations of the null hypotheses (\ref{j0}) and deviations (\ref{jk}), for the significance level $\alpha=0.05$. The rejection rates were computed for the three types of estimation of the functional coefficient and basis representation, in order to see the possible effects of the estimation method in the power performance. At sight of the rejection frequencies for the three models, several comments must be done. Firstly, the test respects the significance levels for the null hypothesis for the three estimation methods considered. Secondly, there seems to be no big differences in terms of power for the three methods, although it can be observed that the FPC and FPLS estimation methods are slightly more conservative. Finally, at sight of the similarities between the response under the null and under the alternatives (see Figure \ref{fig3} in appendix), the results of Table \ref{tab4new} point toward a quite competitive test. Similar results are obtained with a non symmetric random noise. \\

\begin{table}[h]
\centering
\small
\begin{tabular}{c|ccc||ccc}\toprule\toprule
{Models} & B-splines & FPC & FPLS & B-splines & FPC & FPLS\\ \toprule\toprule
$H_{1,0}$ & $0.061$ & $0.052$ & $0.059$ & $0.039$ & $0.046$ & $0.046$ \\
$H_{1,1}$ & $0.094$ & $0.082$ & $0.078$ & $0.074$ & $0.072$ & $0.077$ \\
$H_{1,2}$ & $0.747$ & $0.732$ & $0.715$ & $0.737$ & $0.721$ & $0.720$ \\
$H_{1,3}$ & $0.997$ & $0.997$ & $0.996$ & $0.996$ & $0.997$ & $0.996$ \\\midrule
$H_{2,0}$ & $0.058$ & $0.045$ & $0.050$ & $0.041$ & $0.035$ & $0.033$ \\
$H_{2,1}$ & $0.086$ & $0.071$ & $0.074$ & $0.081$ & $0.080$ & $0.078$ \\
$H_{2,2}$ & $0.745$ & $0.722$ & $0.720$ & $0.743$ & $0.724$ & $0.718$ \\
$H_{2,3}$ & $0.997$ & $0.996$ & $0.997$ & $0.994$ & $0.995$ & $0.994$ \\\midrule
$H_{3,0}$ & $0.054$ & $0.046$ & $0.044$ & $0.052$ & $0.040$ & $0.038$ \\
$H_{3,1}$ & $0.082$ & $0.077$ & $0.075$ & $0.072$ & $0.062$ & $0.062$ \\
$H_{3,2}$ & $0.764$ & $0.752$ & $0.750$ & $0.735$ & $0.737$ & $0.721$ \\
$H_{3,3}$ & $0.999$ & $0.998$ & $0.998$ & $0.998$ & $0.998$ & $0.997$ \\\bottomrule\bottomrule
\end{tabular}
\caption{\small Empirical power of the PCvM test for the composite hypothesis $H_0: m\in \lrb{\inprod{\cdot}{\beta}:\beta\in\Hbb}$ and for three estimating methods of $\beta$ at significance level $\alpha=0.05$ with noise $\mathcal{N}\big(0,0.10^2\big)$ (first three columns) and recentred $\text{Exp}(0.10)$ (last three).\label{tab4new}}
\end{table}

The behaviour of the test for different sample sizes is shown in Table \ref{tab7new}. As in the previous tables, the three estimating methods have very similar rejection ratios and we can see that B-splines estimation has again larger rejection ratios for all the models. As expected, when the sample sizes increases, the rejection rates also do.

\begin{table}[H]
\centering
\small
\begin{widetable}{\columnwidth}{c|ccc|ccc|ccc|ccc}\toprule\toprule
& \multicolumn{3}{c|}{$H_{1,0}$} & \multicolumn{3}{c|}{$H_{1,1}$} & \multicolumn{3}{c|}{$H_{1,2}$} &\multicolumn{3}{c}{$H_{1,3}$} \\
Method & $50$ & $100$ & $200$ & $50$ & $100$ & $200$ & $50$ & $100$ & $200$ & $50$ & $100$ & $200$
\\\midrule
B-spline &  $0.076$ & $0.061$ & $0.062$& $0.093$ & $0.094$ & $0.121$ & $0.484$ & $0.747$ &$0.966$ & $0.900$ & $0.997$ & $1.000$\\
FPC & $0.059$ & $0.052$ & $0.059$ & $0.064$ & $0.082$ & $0.123$ & $0.442$ & $0.732$ & $0.963$ & $0.893$ & $0.997$ & $1.000$ \\
FPLS & $0.062$ & $0.059$ & $0.058$ & $0.069$ & $0.078$ & $0.115$ & $0.414$ & $0.715$ & $0.961$ & $0.873$ & $0.996$ & $1.000$ \\ \bottomrule\bottomrule
\end{widetable}
\caption{\small Empirical power of the PCvM test for the composite hypothesis $H_0: m\in \lrb{\inprod{\cdot}{\beta}:\beta\in\Hbb}$ and for different sample sizes $n$. Noise is a $\mathcal{N}\big(0,0.10^2\big)$. \label{tab7new}}
\end{table}

\section{Data application and graphical tool}
\label{appli}

\begin{figure}[!h]
\centering
\includegraphics[scale=0.3]{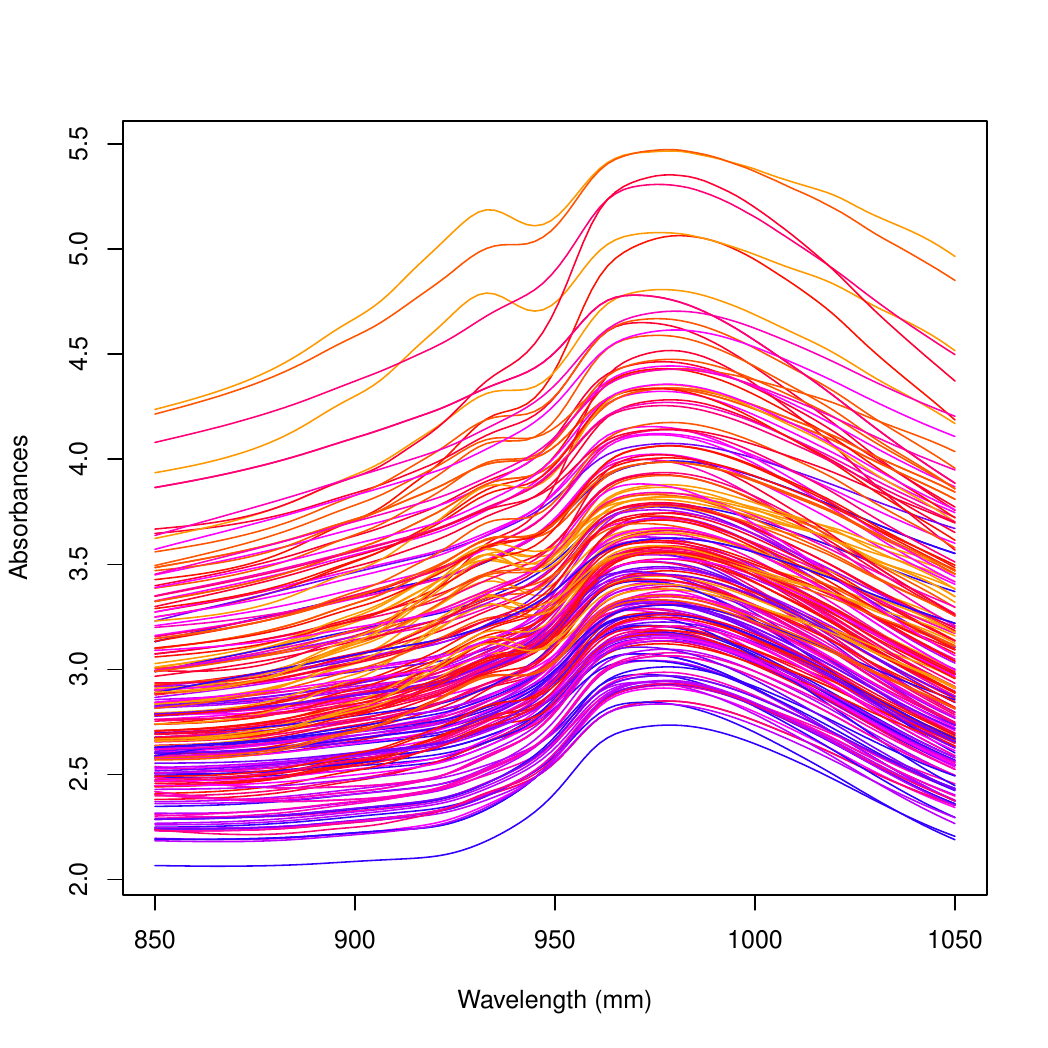}
\centering
\includegraphics[scale=0.3]{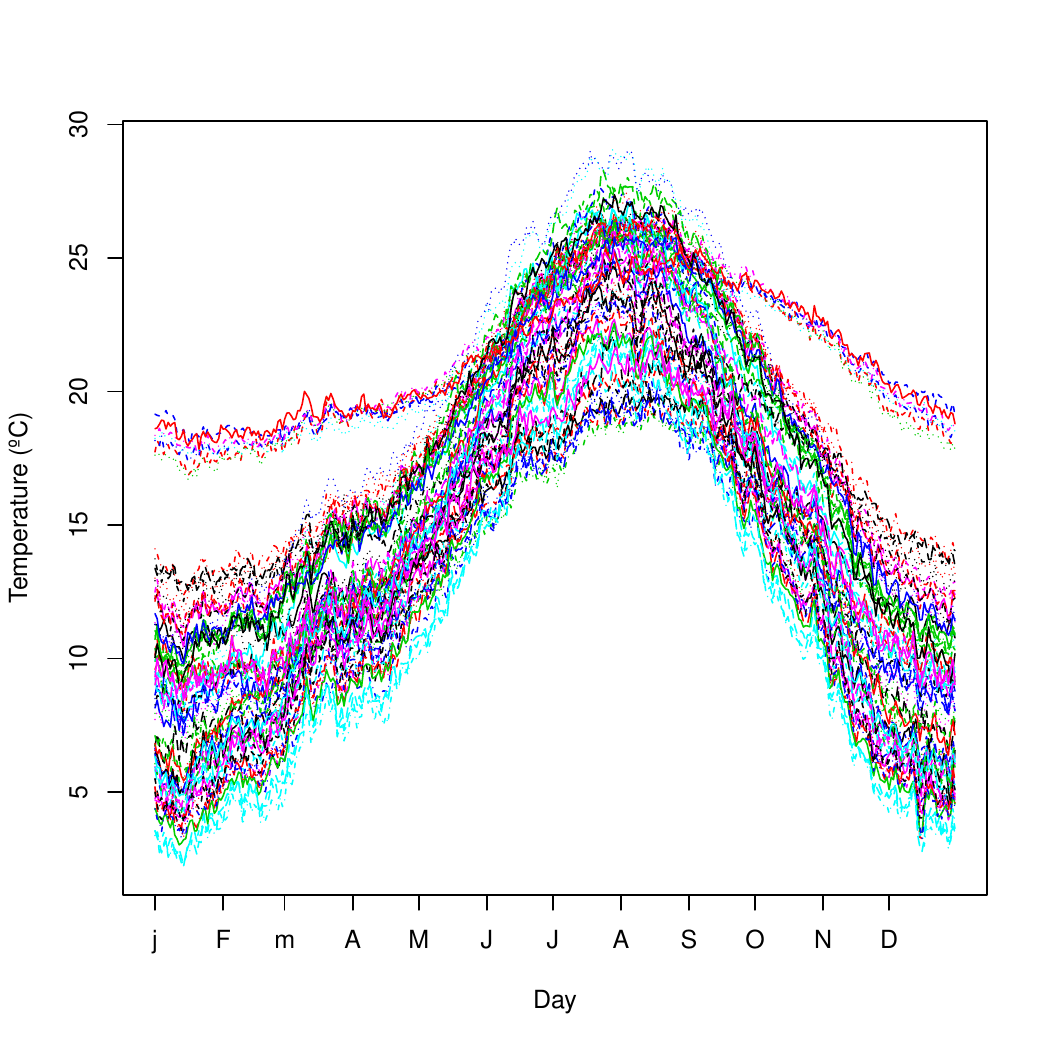}
\centering
\includegraphics[scale=0.3]{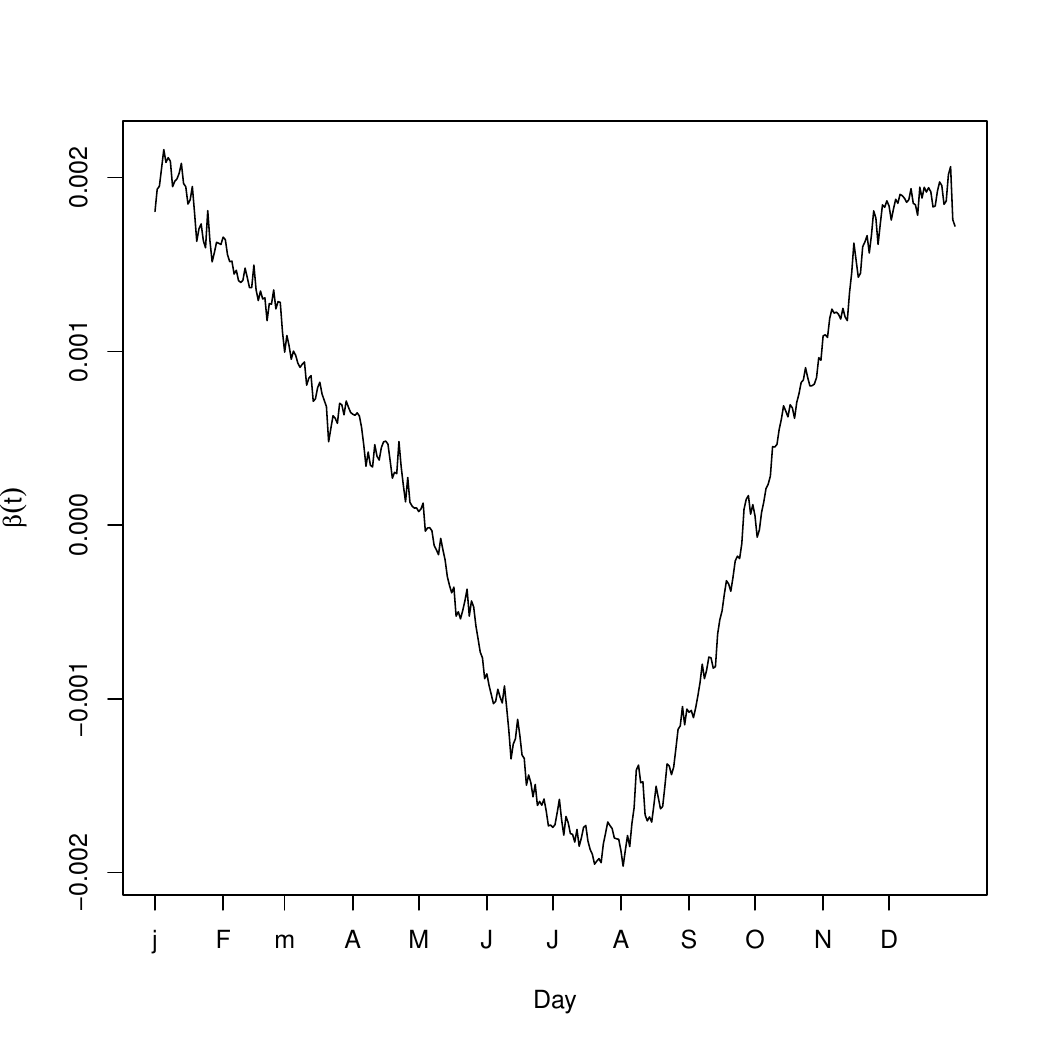}
\caption{\small From left to right: Tecator dataset with spectrometric curves coloured according to their content of fat (red for larger and blue for lower); AEMET temperatures for the $73$ Spanish weather stations; estimated functional coefficient by the FPLS method for the AEMET dataset.\label{fig4}}
\end{figure}

The Tecator dataset is a well known dataset in the literature of functional data analysis (see, for example, \cite{Ferraty2006}). It contains data from $215$ meat samples, consisting of a $100$ channel spectrum of absorbances measured by a spectrometer and the contents of water, fat and protein. When trying to explain the content of fat in the meat samples throughout the spectrometric curves, it is common to transform the original curves into the first derivatives or the second derivatives, in order to properly capture the wavy effects of the meat samples with high percentage of fat (see the left plot of Figure \ref{fig4}).\\

We have applied our goodness-of-fit test with $B=5000$ bootstrap replicates for the original dataset and for the dataset of the first and second derivatives. The $p$-values obtained are $0.004$, $0.000$ and $0.000$, respectively. Thus we have significative evidences against the null hypothesis of FLM. The test was applied with the FPLS estimation method and with automatic selection of the number of FPLS by PCV. As the case of no interaction is a particular case of a FLM, we can conclude that in the Tecator dataset there exists a significative dependence between the functional covariate and the scalar response, although this dependence is not a linear one.\\

The other dataset considered is the AEMET dataset, which is available in the \textbf{R} package \texttt{fda.usc} (see \cite{Febrero-Bande2011}). It is formed by the daily summaries of $73$ Spanish weather stations during the period 1980--2009. Among others, the functional covariate is the daily temperature in each weather station, and the scalar response is the daily wind speed (both variables are averaged over 1980--2009). The center plot of Figure \ref{fig4} represents the functional observations of the daily temperature. Before applying the tests, four functional outliers corresponding to the $5\%$ less depth curves according to the \cite{Fraiman2001} depth were removed.

The resulting $p$-value from the goodness-of-fit test is $0.121$, thus there is no significative evidences to reject the null hypothesis of the FLM for the AEMET dataset. The test is applied with the FPLS estimation method and with $B=5000$ bootstrap replicates. The right plot of Figure \ref{fig4} shows the estimated functional parameter $\beta$, resulting from a basis of $2$ FPLS. Once we have determined that the FLM is a suitable model, we can check if the estimated coefficient $\beta$ is significantly different from zero with the available tests for the simple hypothesis: the functional $F$-test, the test of Delsol et al. (with PCV bandwidth) and our test for the simple null hypothesis of no interaction. The $p$-values obtained are: $0.002$, $0.000$ and $0.000$, respectively. All the tests reject the null, so we can conclude that the curves of the temperature and the average wind speed show a non-trivial linear relation.\\

\begin{figure}[!h]
\centering
\includegraphics[scale=0.425]{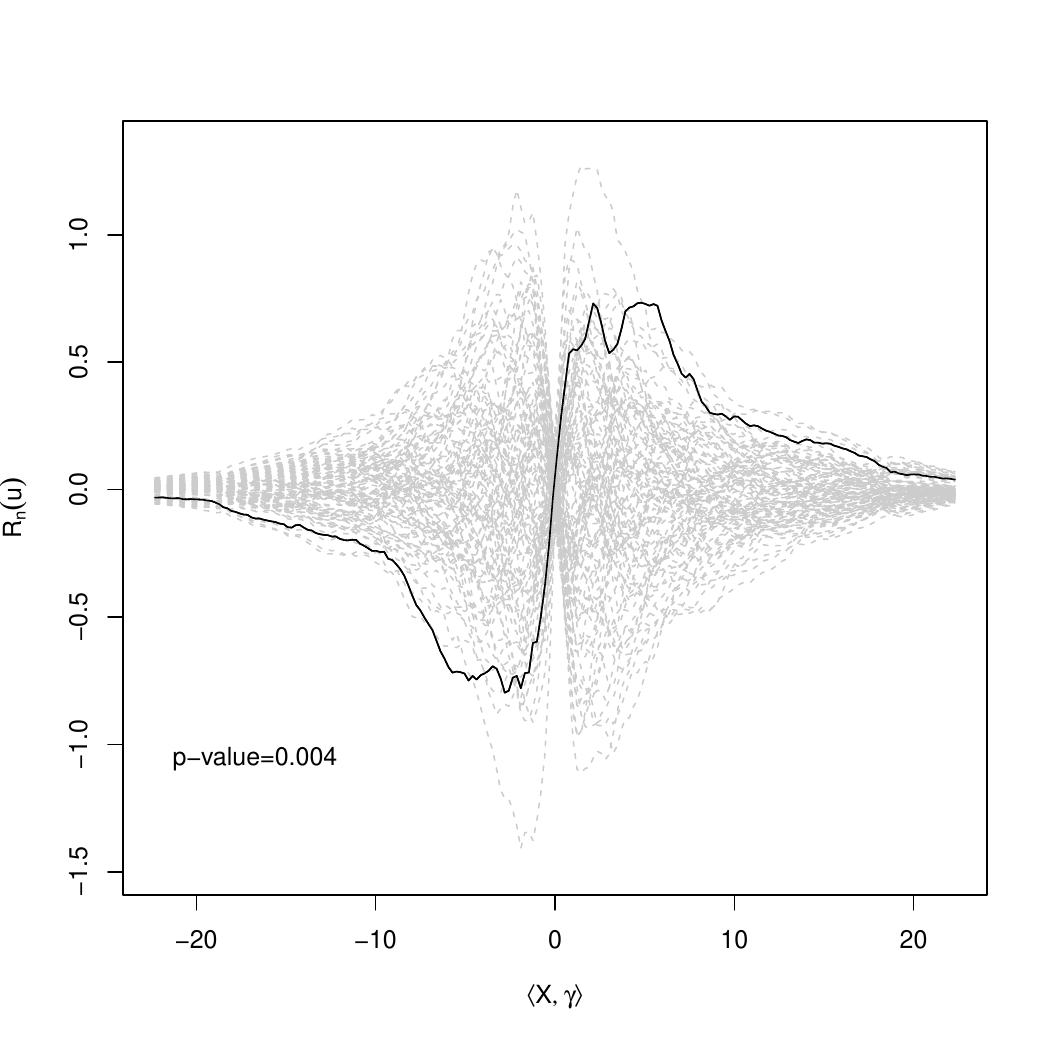}
\centering
\includegraphics[scale=0.425]{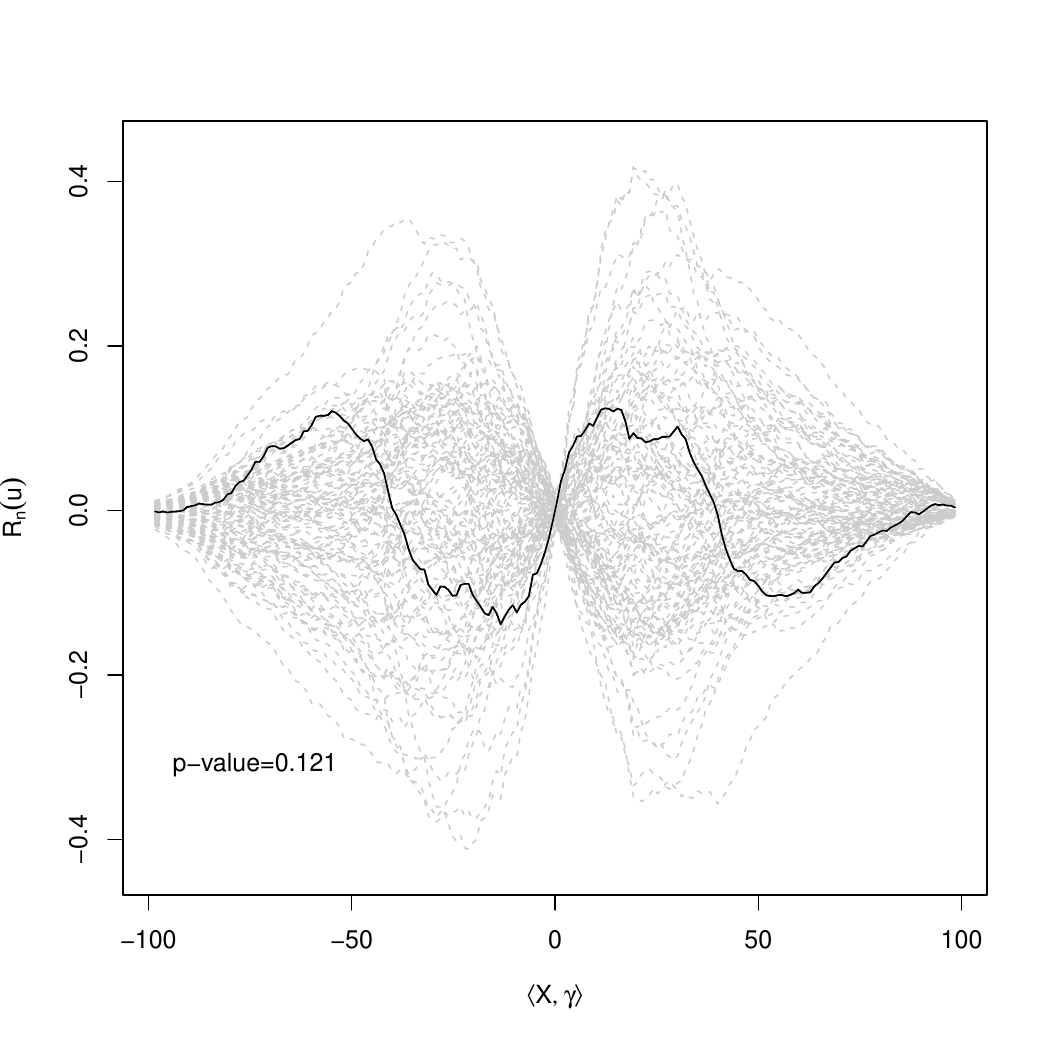}
\caption{\small $R_n$ process observed (solid line) and $B=100$ generated process under the null hypothesis $H_0: m\in \lrb{\inprod{\cdot}{\beta}:\beta\in\Hbb}$ (dashed lines), for the Tecator dataset (left) and the AEMET dataset (right). The number of Monte Carlo replicates for the projections is $G=200$. \label{fig5}}
\end{figure}

We conclude this section showing a graphical tool to visualize the goodness-of-fit of the FLM to a dataset that can be useful to practitioners. The key idea is to compare graphically the process (\ref{rmpp}) obtained with the residuals of the fitted model with the processes obtained with the bootstrapped residuals under the null hypothesis. The path of the RMPP depends on the random projections $\gamma$ and therefore it is difficult to compare two trajectories of the process. However, integrating with respect to $\gamma$ results a process that does not depend on the projections. Further, this integration is easily approximated by Monte Carlo:
\begin{align*}
R_n(u)=\int_{\Sv_\Hbb} R_n(u,\gamma)\,\omega(d\gamma)\approx \frac{1}{G}\sum_{g=1}^GR_n(u,\gamma_g),
\end{align*}
being $\gamma_g$ functions in $\Sv_\Hbb$ and $G$ the number of Monte Carlo replicates. For $\gamma_g$, a possibility is to consider stationary Gaussian processes with unit norm. Figure \ref{fig5} shows the comparison of the observed process $R_n$ and $B=100$ bootstrapped processes under the null, for the two studied datasets. Consistently with the obtained $p$-values, the observed processes for the Tecator dataset seem to be significantly different, whereas for the AEMET dataset the observed process is just an \textit{ordinary} trajectory of the bootstrapped ones.

\section{Conclusions}
\label{final}

We have presented a goodness-of-fit test for the null hypothesis of the functional linear model. The test is constructed adapting the propose of \cite{Escanciano2006} to the functional scheme using a basis representation. Different estimation methods for the functional parameter were considered, showing in general a similar behaviour in the performance of the test. The simulation study shows that the test behaves well in practise: respects the significance level and has good power. The test was applied to two real datasets to determine if the FLM was plausible, rejecting the null hypothesis for the first and finding no evidences for rejecting in the second.\\

The asymptotic distribution of the statistics PCvM$_n$ and PCvM$_{n,p}$, quadratic functionals of the processes $R_n$ and $R_{n,p}$, respectively, is an open problem. The convergence of both processes remains as a problem of great relevance to be considered in the future, taking into account that these processes are indexed in $\R\times\Hbb$ and that it does not exist, up to our knowledge, any results of weak functional convergence  of empirical processes indexed in infinite dimensional spaces.

Although in this paper we have focused on the functional linear model, the proposed test can be extended to checking for any other regression model with functional covariate and scalar response. As the statistic is based on the residuals, the practical implementation and the wild bootstrap calibration given in Section \ref{thetest} will remain the same: we just have to consider suitable estimators for the parameters of the regression model to compute the residuals. Therefore, obvious extensions could be the testing of FLM with several covariates or the testing of the quadratic functional model.\\

Finally, let us remark that the code for the implementation of the goodness-of-fit test in the simple and composite cases is available throughout the function \texttt{flm.test} of the \texttt{R} library \texttt{fda.usc} since version 0.9.8. This function also shows the graphical tool introduced in Section \ref{appli}. To speed up the computation of the test statistic, the critical parts of the test implementation have been programmed in \texttt{FORTRAN}.

\section*{Supplementary materials}

The supplementary materials contain the proof of Lemma \ref{p1}, and figures and more detailed tables for the results of the simulation study.

\section*{Acknowledgements}

The authors acknowledge the support of Project MTM2008-03010, from the Spanish Ministry of Science and Innovation, Project 10MDS207015PR from Direcci\'on Xeral de I+D, Xunta de Galicia  and IAP network StUDyS, from Belgian Science Policy. Work of E. Garc\'ia-Portugu\'es has been supported by FPU grant AP2010-0957 from the Spanish Ministry of Education. The authors also acknowledge the suggestions by two anonymous referees that helped improving this paper.

\appendix

\newpage

\title{Supplementary material for ``A goodness-of-fit test for the functional linear model with scalar response''}
\setlength{\droptitle}{-1cm}
\predate{}%
\postdate{}%
\date{}

\author{Eduardo Garc\'ia-Portugu\'es$^{1,2}$, Wenceslao Gonz\'alez-Manteiga$^{1}$, and Manuel Febrero-Bande$^1$}

\footnotetext[1]{
Department of Statistics and Operations Research, University of Santiago de Compostela (Spain).}
\footnotetext[2]{Corresponding author. e-mail: \href{mailto:eduardo.garcia@usc.es}{eduardo.garcia@usc.es}.}

\maketitle

\begin{abstract}
This supplementary material is formed by three sections. The first one contains the proof of Lemma \ref{p1}, while the second and third show figures and tables omitted from the paper that add more insights on the simulation study.
\end{abstract}

\begin{flushleft}
\small
\textbf{Keywords:} Bootstrap calibration; Functional data; Functional linear model; Goodness-of-fit.
\end{flushleft}

\section{Proof of Lemma \ref{p1}}

\begin{proof}[Proof of Lemma \ref{p1}]
Let $\beta$ be an element of $\Hbb$. We will proceed by proving equivalences by pairs. \\

First of all, equivalence of \ref{p1:1} and \ref{p1:2} is immediately  by the definition of $m(x)=\espe{Y|\Xcal=x}$. Equivalences of \ref{p1:2}, \ref{p1:3} and \ref{p1:3b} follow by Lemma \ref{l1}.\\

The equivalence of \ref{p1:3} and \ref{p1:4} is based on the definition of the integrated regression function and is given by a chain of equivalences. Let denote $U_\gamma=\inprod{X}{\gamma}$, for any $\gamma\in\Sv_\Hbb$, $m_\gamma(u)=\espe{Y|U_\gamma=u}$ and $m_{0,\gamma}(u)=\espe{\inprod{\Xcal}{\beta}|U_\gamma=u}$. The integrated regression functions for $m_\gamma$ and $m_{0,\gamma}$ are given by:
\begin{align}
I_\gamma(u)&=\espe{Y\ind{U_\gamma\leq u}}=\espe{\espe{Y\ind{U_\gamma\leq u}|U_\gamma}}=\espe{\espe{Y|U_\gamma}\ind{U_\gamma\leq u}}\nonumber\\
&=\espe{m_\gamma(U_\gamma)\ind{U_\gamma\leq u}}=\int_{-\infty}^\infty m_\gamma(u)\ind{u\leq x}\,dF_\gamma(u)=\int_{-\infty}^x m_\gamma(u)\,dF_\gamma(u),\label{p1:p:1}\\
I_{0,\gamma}(u)&=\espe{\inprod{\Xcal}{\beta}\ind{U_\gamma\leq u}}=\espe{\espe{\inprod{\Xcal}{\beta}\ind{U_\gamma\leq u}|U_\gamma}}=\espe{\espe{\inprod{\Xcal}{\beta}|U_\gamma}\ind{U_\gamma\leq u}}\nonumber\\
&=\espe{m_{0,\gamma}(U_\gamma)\ind{U_\gamma\leq u}}=\int_{-\infty}^\infty m_{0,\gamma}(u)\ind{u\leq x}\,dF_\gamma(u)=\int_{-\infty}^x m_{0,\gamma}(u)\,dF_\gamma(u),\label{p1:p:2}
\end{align}
where $F_\gamma$ represents the distribution function of $U_\gamma$. Statement \ref{p1:3} can be expressed as
\begin{align*}
m_{\gamma}(u)=m_{0,\gamma}(u),\text{ for a.e. }u\in\RR,
\end{align*}
which by (\ref{p1:p:1}) and (\ref{p1:p:2}) is equivalent to 
\begin{align}
I_{\gamma}(u)=I_{0,\gamma}(u),\text{ for a.e. }u\in\RR. \label{las}
\end{align}
As \ref{p1:4} is equivalent to (\ref{las}), this proofs the equivalence of \ref{p1:3} and \ref{p1:4}. The same argument can be applied to prove the equivalence between \ref{p1:3b} and \ref{p1:4b}, which ends the proof.
\end{proof}

\section{Figures}


\vspace*{\fill}

\begin{figure}[H]
\centering
\includegraphics[scale=0.5]{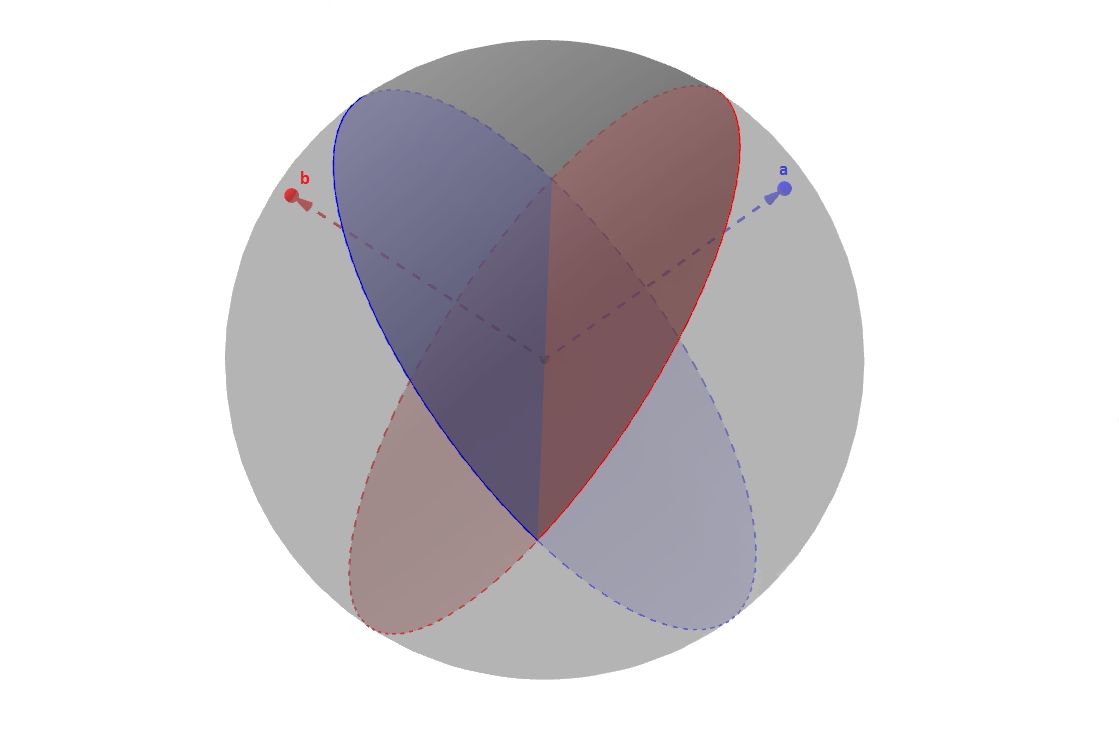}
\caption{\small Spherical wedge $S_{\mathbf{a},\mathbf{b}}=\lrb{\boldsymbol\xi\in\Sv^p: \frac{\pi}{2}\leq \measuredangle\lrp{\boldsymbol\xi,\mathbf{a}}\leq\frac{3\pi}{2},\,\frac{\pi}{2}\leq \measuredangle\lrp{\boldsymbol\xi,\mathbf{b}}\leq\frac{3\pi}{2}}$ defined by points $\mathbf{a}$ and $\mathbf{b}$ in $\Sv^3=\lrb{\mathbf x\in\R^3:\norm{\mathbf x}=1}$. The wedge is the region of the sphere determined by the intersection of the subspaces that are generated by the normal planes of the vectors $\mathbf{a}$ and $\mathbf{b}$.\label{fig0}}
\end{figure}
\vspace*{\fill}

\pagebreak

\vspace*{\fill}

\begin{figure}[H]
\centering
\includegraphics[scale=0.45]{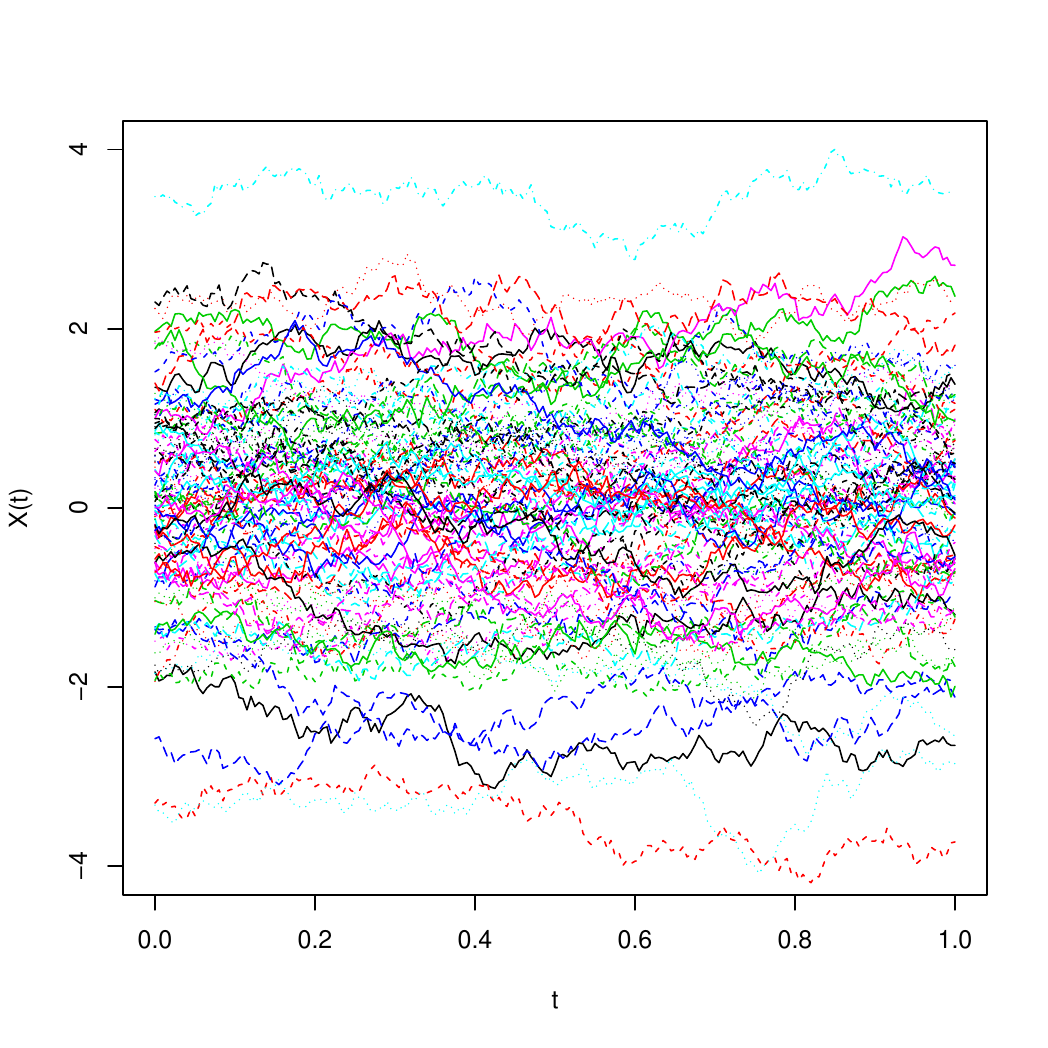}
\centering
\includegraphics[scale=0.45]{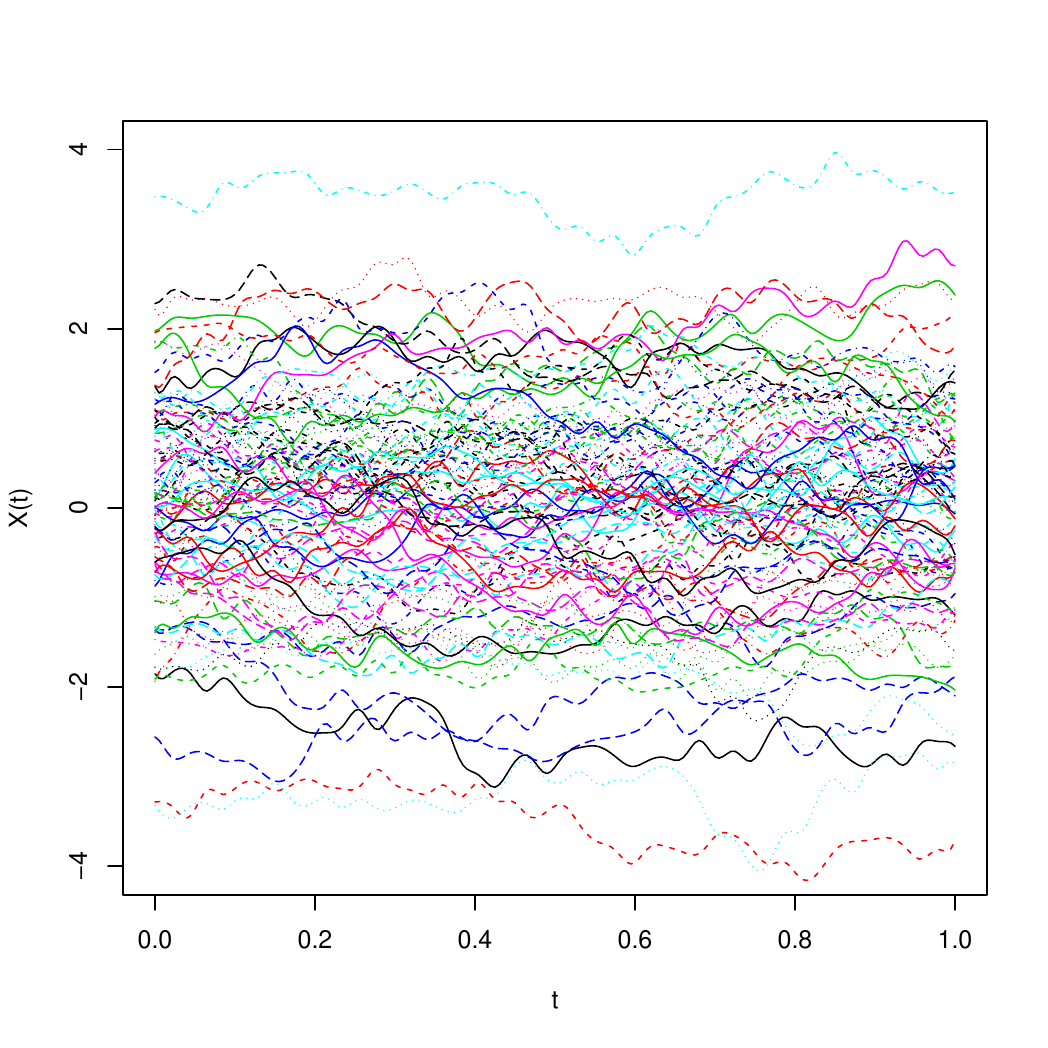}
\centering
\includegraphics[scale=0.45]{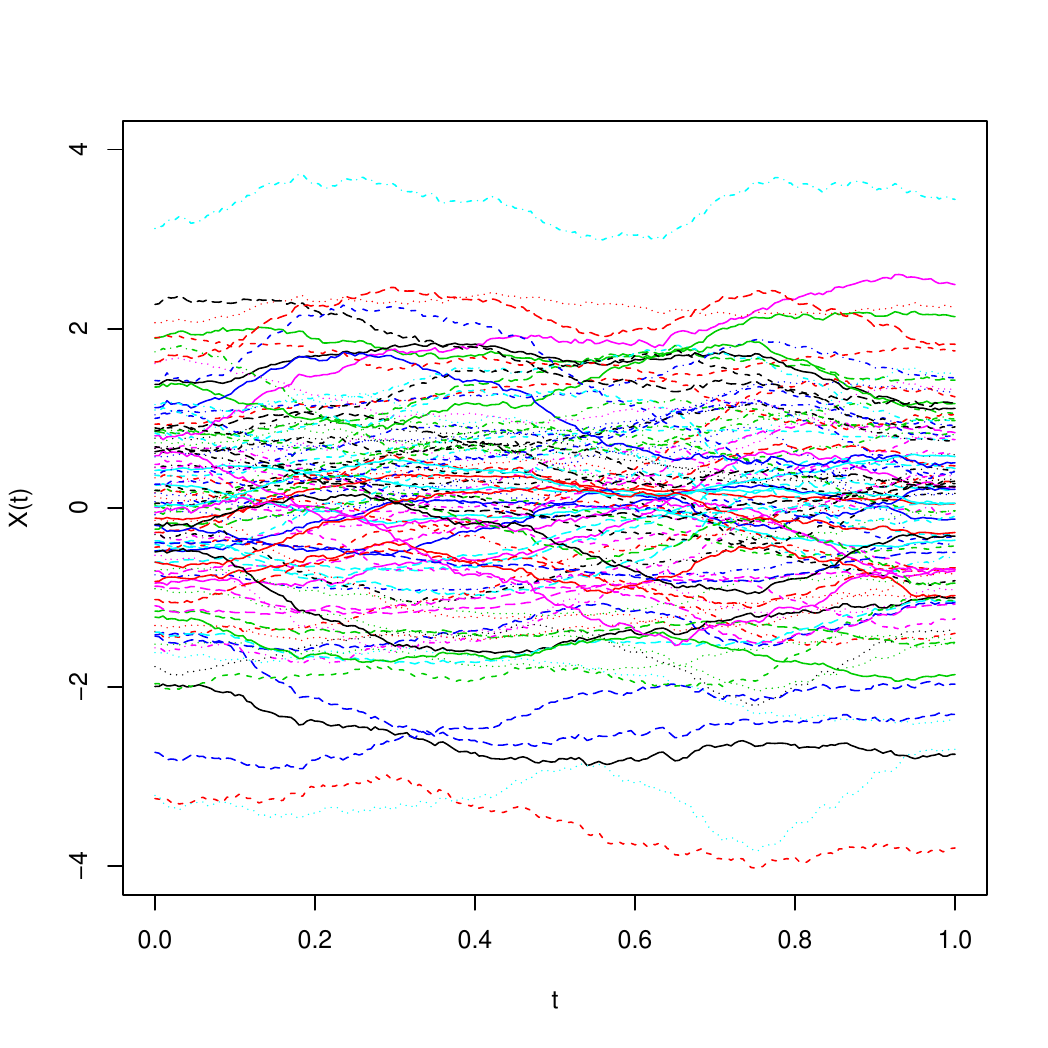}
\centering
\includegraphics[scale=0.45]{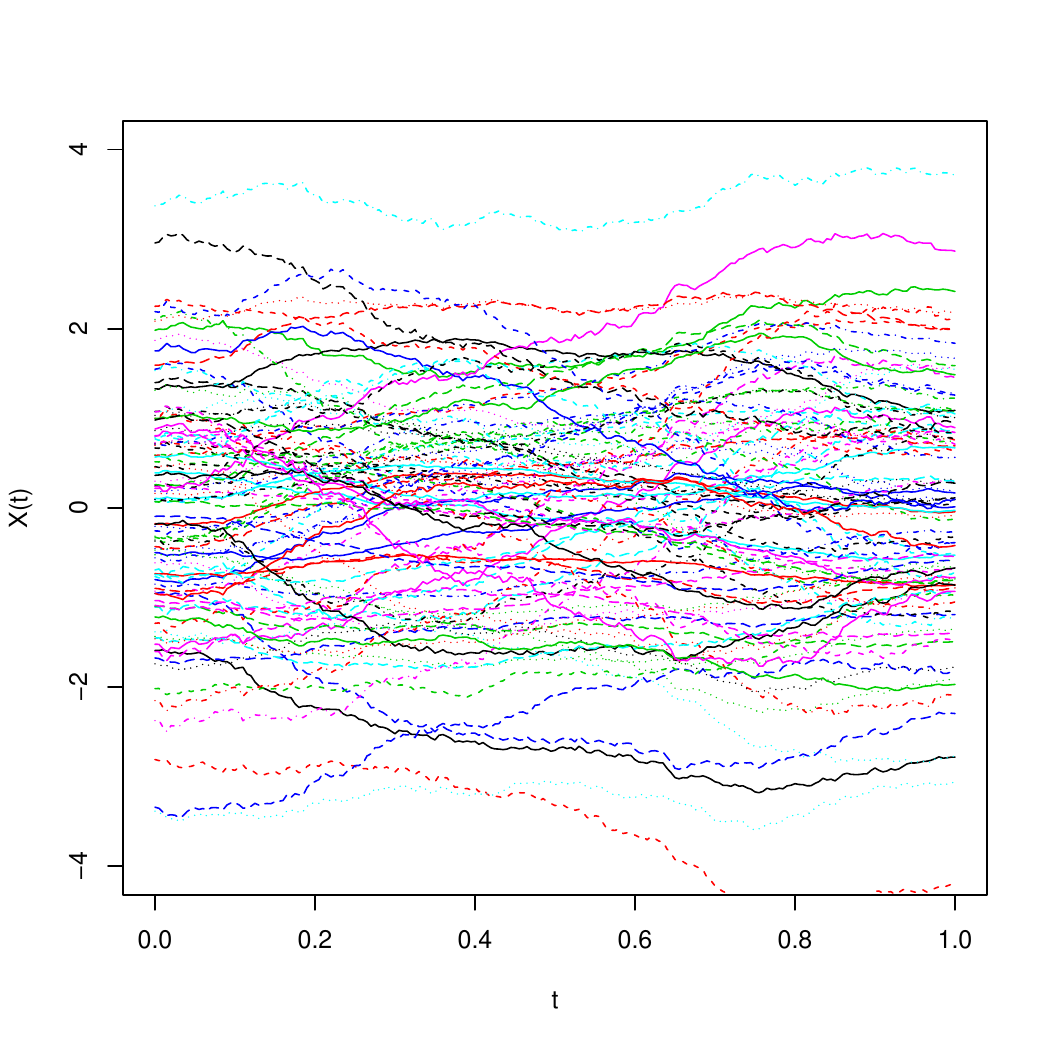}
\caption{\small The effect of the kind and length of basis expansion for functional data. From up to down and left to right: sample of $100$ simulated trajectories from the Ornstein--Uhlenbeck process; representation in a B-splines basis of $50$ elements; representation in a FPC basis of $5$ elements; representation in a FPLS basis of $5$ elements, using an independent scalar response distributed as a $\mathcal{N}(0,1)$. \label{fig1}}
\end{figure}

\vspace*{\fill}

\pagebreak

\vspace*{\fill}

\begin{figure}[H]
\centering
\includegraphics[scale=0.4]{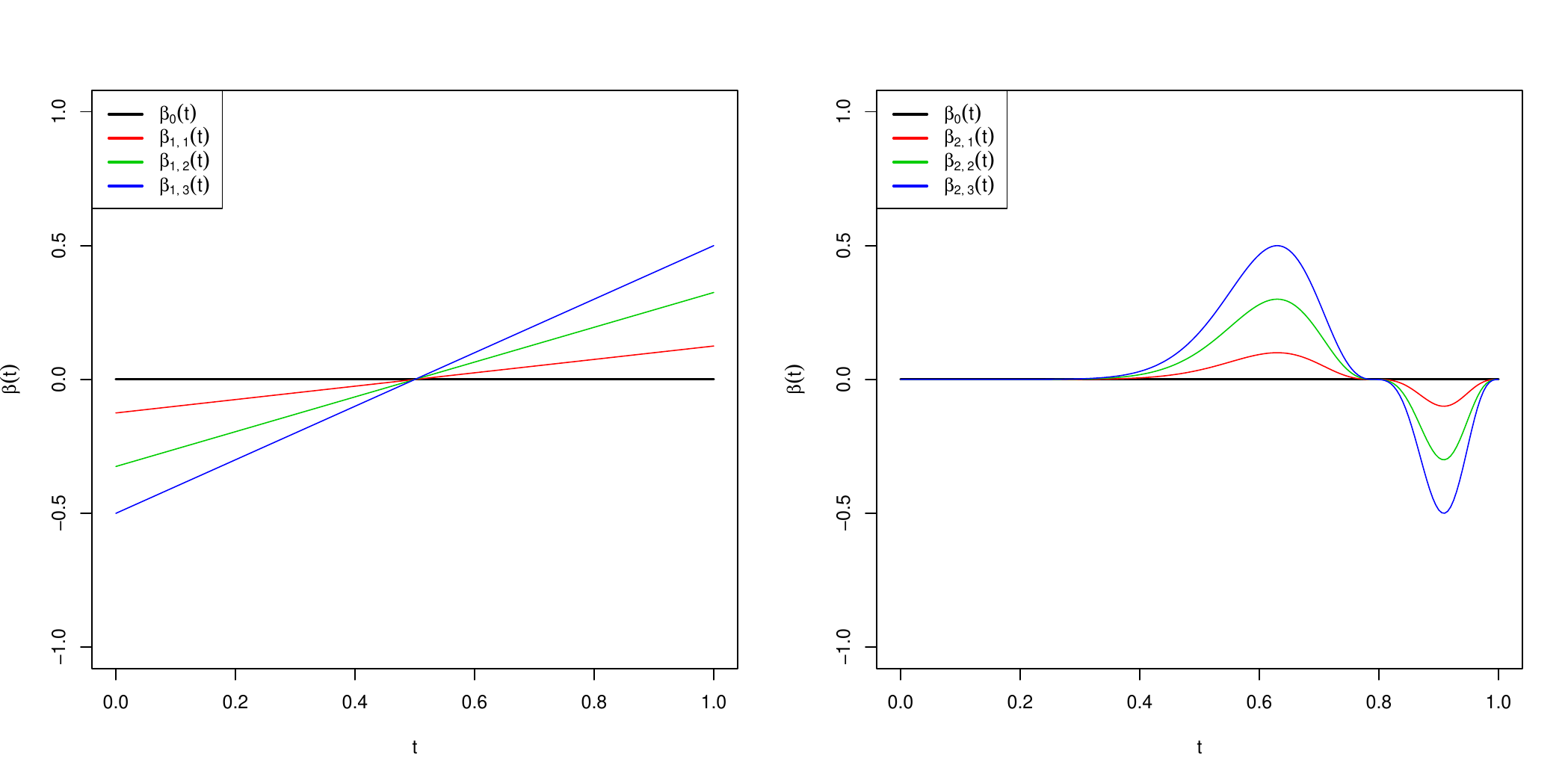}
\centering
\includegraphics[scale=0.375]{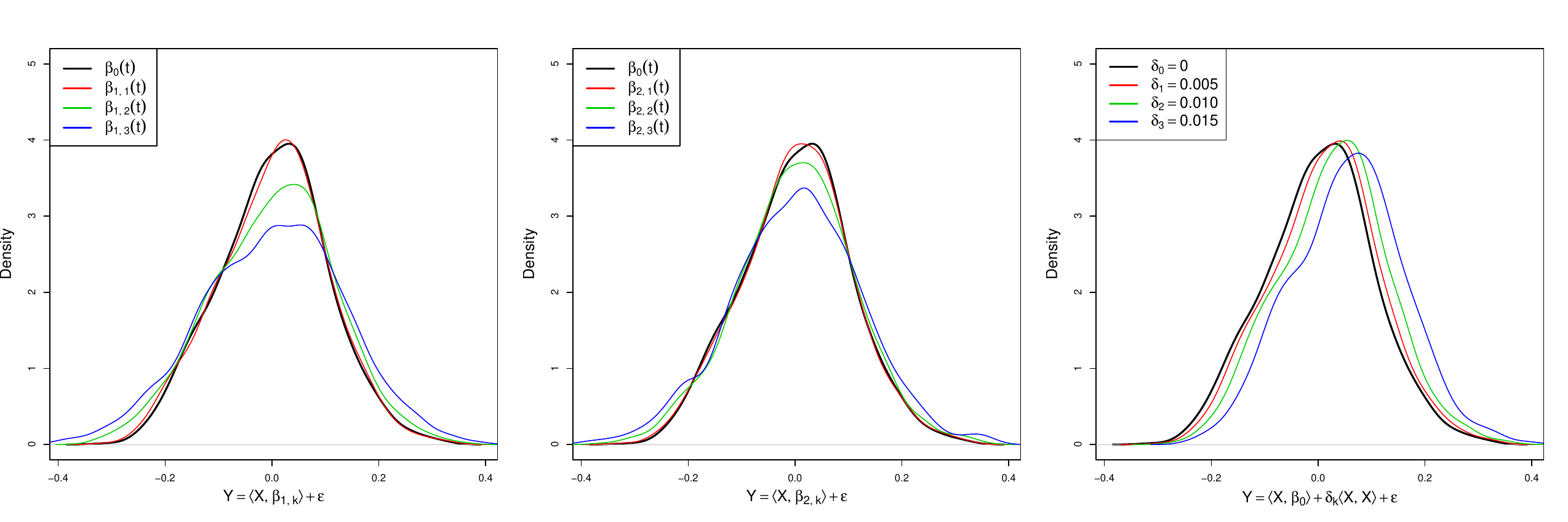}
\caption{\small Upper row: functional coefficient deviations of the simple null hypothesis for $H_{1,k}$ (left) and $H_{2,k}$ (right), $k=1,2,3$. Lower row: densities of the scalar response under the null hypothesis ($H_0$) and for the three deviations ($H_{j,k},\,k=1,2,3,$ for each model $j=1,2,3$). The estimation of the densities of the response has been done with kernel smoothing from a sample of $1000$ observations. The bandwidth is the same in the four densities of each model, and is computed by the method of \cite{Sheather1991}, for the case of the null hypothesis. \label{fig2}}
\end{figure}

\vspace*{\fill}

\pagebreak

\vspace*{\fill}

\begin{figure}[H]
\centering
\includegraphics[scale=0.375]{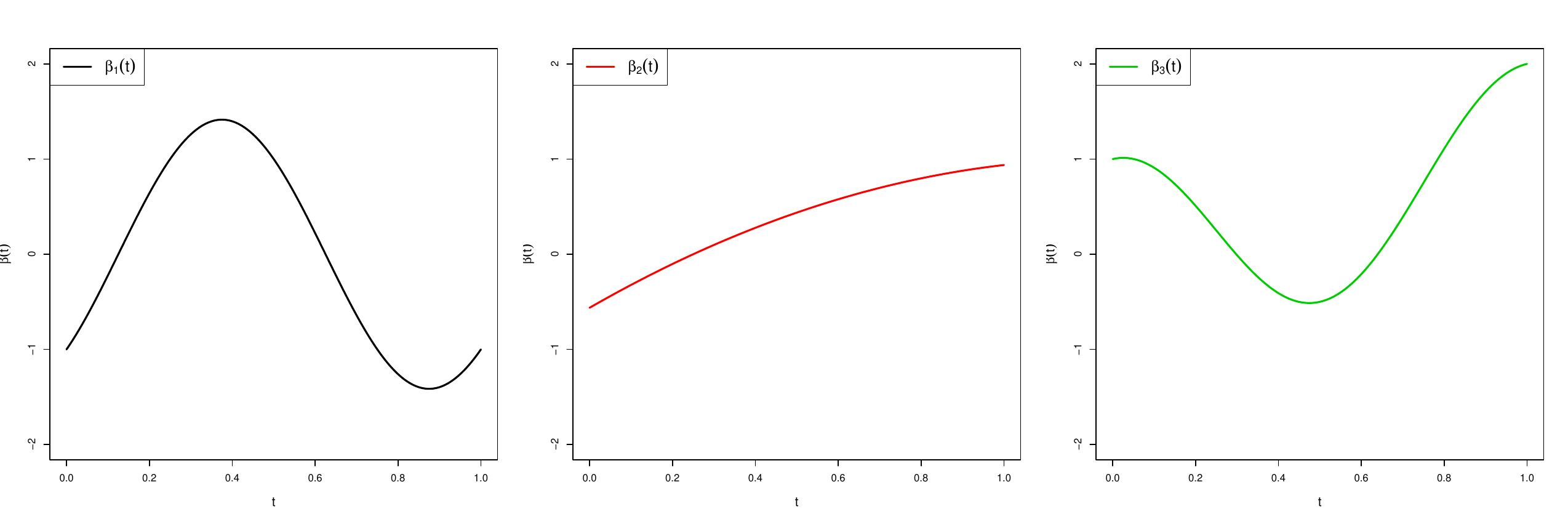}
\centering
\includegraphics[scale=0.375]{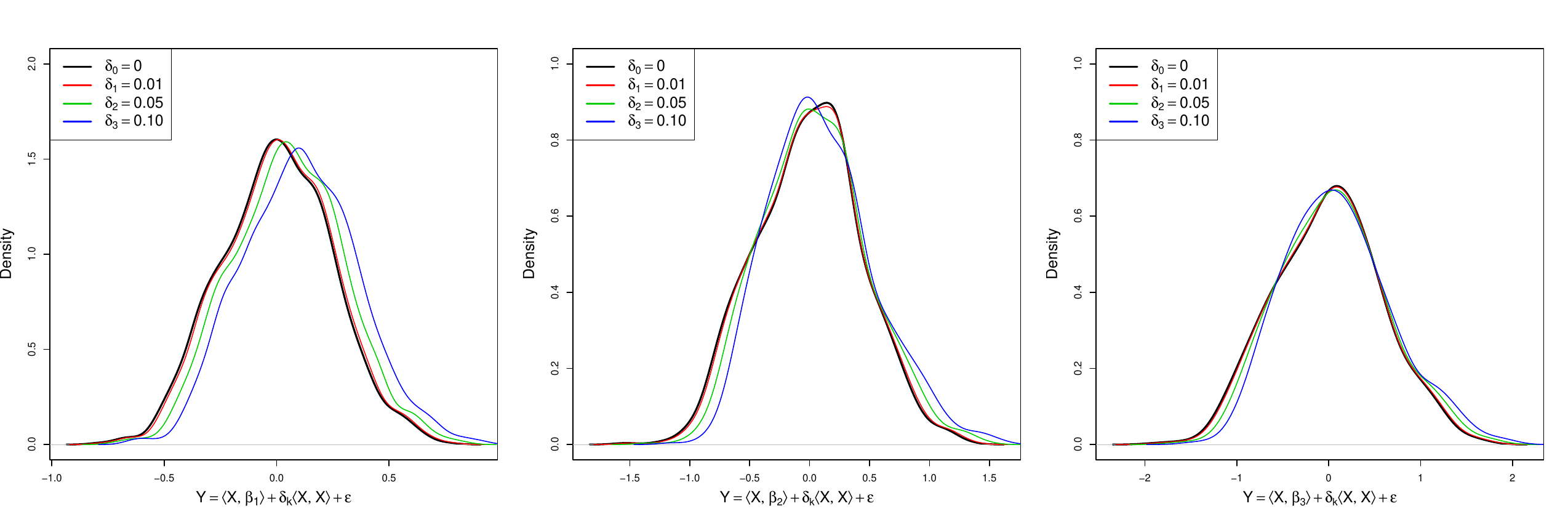}
\caption{\small Upper row: functional coefficients of the linear models for the composite hypothesis. Lower row: densities of the scalar response under the null hypothesis ($H_{j,0},$ for each model $j=1,2,3$) and for the three quadratic deviations ($H_{j,k},\,k=1,2,3,$ for each model $j=1,2,3$). The densities are computed in the way described for the simple hypothesis. \label{fig3}}
\end{figure}

\vspace*{\fill}

\pagebreak

\section{Tables}


\subsection{Simple hypothesis}

\begin{table}[!h]
\centering
\small
\begin{tabular}{c|c|c|ccccc}\toprule\toprule
\multirow{2}{*}{Models} & \multirow{2}{*}{$F$-test} & \multirow{2}{*}{PCvM test} & \multicolumn{5}{c}{Delsol {et al.} test}\\
& & & $h=h_{CV}$ & $h=0.25$ & $h=0.50$ & $h=0.75$ & $h=1.00$ \\\midrule
$H_0$ & $0.060$ & $0.041$ & $0.065$ & $0.043$ & $0.055$ & $0.050$ & $0.047$ \\\midrule
$H_{1,1}$ & $0.060$ & $0.069$ & $0.098$ & $0.069$ & $0.070$ & $0.069$ & $0.068$ \\
$H_{1,2}$ & $0.163$ & $0.078$ & $0.309$ & $0.350$ & $0.112$ & $0.063$ & $0.057$ \\
$H_{1,3}$ & $0.401$ & $0.138$ & $0.772$ & $0.815$ & $0.241$ & $0.087$ & $0.066$ \\\midrule
$H_{2,1}$ & $0.248$ & $0.053$ & $0.080$ & $0.068$ & $0.078$ & $0.066$ & $0.055$ \\
$H_{2,2}$ & $0.951$ & $0.336$ & $0.403$ & $0.318$ & $0.447$ & $0.391$ & $0.274$ \\
$H_{2,3}$ & $1.000$ & $0.904$ & $0.877$ & $0.794$ & $0.887$ & $0.870$ & $0.775$ \\\midrule
$H_{3,1}$ & $0.034$ & $0.173$ & $0.165$ & $0.051$ & $0.096$ & $0.116$ & $0.126$ \\
$H_{3,2}$ & $0.038$ & $0.691$ & $0.554$ & $0.209$ & $0.361$ & $0.456$ & $0.522$ \\
$H_{3,3}$ & $0.019$ & $0.998$ & $0.932$ & $0.799$ & $0.926$ & $0.956$ & $0.976$ \\\bottomrule\bottomrule
\end{tabular}
\caption{\small Empirical power of the competing tests for the simple hypothesis $H_0:m(\Xcal)=\inprod{\Xcal}{\beta_0}$, $\beta_0(t)=0,\, \forall t$ and significance level $\alpha=0.05$. Noise follows a $\mathcal{N}\big(0,0.10^2\big)$.\label{tab1b}}
\end{table}

\begin{table}[H]
\centering
\small
\begin{tabular}{c|c|c|ccccc}\toprule\toprule
\multirow{2}{*}{Models} & \multirow{2}{*}{$F$-test} & \multirow{2}{*}{PCvM test} & \multicolumn{5}{c}{Delsol {et al.} test}\\
& & & $h=h_{CV}$ & $h=0.25$ & $h=0.50$ & $h=0.75$ & $h=1.00$ \\\midrule
$H_0$ & $0.043$ & $0.051$ & $0.066$ & $0.034$ & $0.054$ & $0.057$ & $0.057$ \\\midrule
$H_{1,1}$ & $0.056$ & $0.052$ & $0.072$ & $0.051$ & $0.055$ & $0.049$ & $0.052$ \\
$H_{1,2}$ & $0.180$ & $0.085$ & $0.285$ & $0.333$ & $0.132$ & $0.065$ & $0.055$ \\
$H_{1,3}$ & $0.442$ & $0.166$ & $0.719$ & $0.773$ & $0.260$ & $0.099$ & $0.074$ \\\midrule
$H_{2,1}$ & $0.265$ & $0.071$ & $0.089$ & $0.052$ & $0.092$ & $0.074$ & $0.071$ \\
$H_{2,2}$ & $0.932$ & $0.343$ & $0.420$ & $0.314$ & $0.460$ & $0.409$ & $0.306$ \\
$H_{2,3}$ & $0.999$ & $0.901$ & $0.848$ & $0.745$ & $0.874$ & $0.856$ & $0.775$ \\\midrule
$H_{3,1}$ & $0.052$ & $0.125$ & $0.128$ & $0.036$ & $0.066$ & $0.077$ & $0.093$ \\
$H_{3,2}$ & $0.034$ & $0.721$ & $0.558$ & $0.136$ & $0.347$ & $0.444$ & $0.526$ \\
$H_{3,3}$ & $0.012$ & $1.000$ & $0.967$ & $0.805$ & $0.985$ & $0.993$ & $0.994$ \\\bottomrule\bottomrule
\end{tabular}
\caption{\small Empirical power of the competing tests for the simple hypothesis $H_0:m(\Xcal)=\inprod{\Xcal}{\beta_0}$, $\beta_0(t)=0,\, \forall t$ and significance level $\alpha=0.05$. Noise follows a recentred $\text{Exp}(10)$.\label{tab2}}
\end{table}

\subsection{Composite hypothesis}

\begin{table}[H]
\centering
\small
\begin{tabular}{c|ccc|ccc|ccc}\toprule\toprule
\multicolumn{1}{c}{}& \multicolumn{9}{c}{Coefficient estimation}\\\midrule
\multirow{2}{*}{Models} & \multicolumn{3}{c|}{B-splines estimation} & \multicolumn{3}{c|}{FPC estimation} & \multicolumn{3}{c}{FPLS estimation}\\
& \footnotesize$\alpha=0.10$ & \footnotesize$\alpha=0.05$ & \footnotesize$\alpha=0.01$ & \footnotesize$\alpha=0.10$ & \footnotesize$\alpha=0.05$ & \footnotesize$\alpha=0.01$ & \footnotesize$\alpha=0.10$ & \footnotesize$\alpha=0.05$ & \footnotesize$\alpha=0.01$\\\midrule
$H_{1,0}$ & $0.125$ & $0.061$ & $0.014$ & $0.107$ & $0.052$ & $0.011$ & $0.100$ & $0.059$ & $0.016$ \\
$H_{1,1}$ & $0.162$ & $0.094$ & $0.025$ & $0.143$ & $0.082$ & $0.025$ & $0.153$ & $0.078$ & $0.021$ \\
$H_{1,2}$ & $0.839$ & $0.747$ & $0.509$ & $0.826$ & $0.732$ & $0.487$ & $0.810$ & $0.715$ & $0.470$ \\
$H_{1,3}$ & $1.000$ & $0.997$ & $0.986$ & $1.000$ & $0.997$ & $0.982$ & $1.000$ & $0.996$ & $0.974$ \\\midrule
$H_{2,0}$ & $0.119$ & $0.058$ & $0.016$ & $0.090$ & $0.045$ & $0.012$ & $0.091$ & $0.050$ & $0.014$ \\
$H_{2,1}$ & $0.164$ & $0.086$ & $0.021$ & $0.149$ & $0.071$ & $0.020$ & $0.154$ & $0.074$ & $0.020$ \\
$H_{2,2}$ & $0.844$ & $0.745$ & $0.513$ & $0.817$ & $0.722$ & $0.491$ & $0.813$ & $0.720$ & $0.489$ \\
$H_{2,3}$ & $0.997$ & $0.997$ & $0.983$ & $0.999$ & $0.996$ & $0.985$ & $0.999$ & $0.997$ & $0.984$ \\\midrule
$H_{3,0}$ & $0.113$ & $0.054$ & $0.009$ & $0.098$ & $0.046$ & $0.008$ & $0.101$ & $0.044$ & $0.008$ \\
$H_{3,1}$ & $0.157$ & $0.082$ & $0.016$ & $0.153$ & $0.077$ & $0.012$ & $0.145$ & $0.075$ & $0.013$ \\
$H_{3,2}$ & $0.853$ & $0.764$ & $0.515$ & $0.838$ & $0.752$ & $0.506$ & $0.834$ & $0.750$ & $0.478$ \\
$H_{3,3}$ & $0.999$ & $0.999$ & $0.986$ & $0.999$ & $0.998$ & $0.987$ & $0.999$ & $0.998$ & $0.985$ \\\bottomrule\bottomrule
\end{tabular}
\caption{\small Empirical power of the PCvM test for the composite hypothesis $H_0: m\in \lrb{\inprod{\cdot}{\beta}:\beta\in\Hbb}$ and for three estimating methods of $\beta$. Noise follows a $\mathcal{N}\big(0,0.10^2\big)$.\label{tab4}}
\end{table}

\begin{table}[H]
\centering
\small
\begin{tabular}{c|ccc|ccc|ccc}\toprule\toprule
\multicolumn{1}{c}{}& \multicolumn{9}{c}{Coefficient estimation}\\\midrule
\multirow{2}{*}{Models} & \multicolumn{3}{c|}{B-splines estimation} & \multicolumn{3}{c|}{FPC estimation} & \multicolumn{3}{c}{FPLS estimation}\\
& \footnotesize$\alpha=0.10$ & \footnotesize$\alpha=0.05$ & \footnotesize$\alpha=0.01$ & \footnotesize$\alpha=0.10$ & \footnotesize$\alpha=0.05$ & \footnotesize$\alpha=0.01$ & \footnotesize$\alpha=0.10$ & \footnotesize$\alpha=0.05$ & \footnotesize$\alpha=0.01$\\\midrule
$H_{1,0}$ & $0.105$ & $0.039$ & $0.004$ & $0.091$ & $0.046$ & $0.005$ & $0.100$ & $0.046$ & $0.006$ \\
$H_{1,1}$ & $0.149$ & $0.074$ & $0.020$ & $0.134$ & $0.072$ & $0.015$ & $0.146$ & $0.077$ & $0.019$ \\
$H_{1,2}$ & $0.823$ & $0.737$ & $0.500$ & $0.813$ & $0.721$ & $0.480$ & $0.801$ & $0.720$ & $0.477$ \\
$H_{1,3}$ & $0.998$ & $0.996$ & $0.986$ & $0.999$ & $0.997$ & $0.987$ & $0.996$ & $0.996$ & $0.983$ \\\midrule
$H_{2,0}$ & $0.089$ & $0.041$ & $0.009$ & $0.087$ & $0.035$ & $0.009$ & $0.088$ & $0.033$ & $0.010$ \\
$H_{2,1}$ & $0.160$ & $0.081$ & $0.020$ & $0.146$ & $0.080$ & $0.016$ & $0.133$ & $0.078$ & $0.015$ \\
$H_{2,2}$ & $0.835$ & $0.743$ & $0.487$ & $0.811$ & $0.724$ & $0.489$ & $0.809$ & $0.718$ & $0.493$ \\
$H_{2,3}$ & $0.995$ & $0.994$ & $0.978$ & $0.996$ & $0.995$ & $0.979$ & $0.995$ & $0.994$ & $0.978$ \\\midrule
$H_{3,0}$ & $0.104$ & $0.052$ & $0.006$ & $0.089$ & $0.040$ & $0.005$ & $0.087$ & $0.038$ & $0.004$ \\
$H_{3,1}$ & $0.130$ & $0.072$ & $0.017$ & $0.119$ & $0.062$ & $0.014$ & $0.110$ & $0.062$ & $0.014$ \\
$H_{3,2}$ & $0.831$ & $0.735$ & $0.498$ & $0.833$ & $0.737$ & $0.486$ & $0.820$ & $0.721$ & $0.481$ \\
$H_{3,3}$ & $0.999$ & $0.998$ & $0.987$ & $0.999$ & $0.998$ & $0.988$ & $0.999$ & $0.997$ & $0.984$ \\\bottomrule\bottomrule
\end{tabular}
\caption{\small Empirical power of the PCvM test for the composite hypothesis $H_0: m\in \lrb{\inprod{\cdot}{\beta}:\beta\in\Hbb}$ and for three estimating methods of $\beta$. Noise follows a recentred $\text{Exp}(10)$.\label{tab5}}
\end{table}

\begin{table}[H]
\centering
\small
\begin{tabular}{cc|ccc|ccc|ccc}\toprule\toprule
\multicolumn{2}{c}{}& \multicolumn{9}{c}{Coefficient estimation}\\\midrule
\multirow{2}{*}{Models} & \multirow{2}{*}{$n$} & \multicolumn{3}{c|}{B-splines estimation} & \multicolumn{3}{c|}{FPC estimation} & \multicolumn{3}{c}{FPLS estimation}\\
 & & \footnotesize$\alpha=0.10$ & \footnotesize$\alpha=0.05$ & \footnotesize$\alpha=0.01$ & \footnotesize$\alpha=0.10$ & \footnotesize$\alpha=0.05$ & \footnotesize$\alpha=0.01$ & \footnotesize$\alpha=0.10$ & \footnotesize$\alpha=0.05$ & \footnotesize$\alpha=0.01$\\\midrule
$H_{1,0}$ & $50$ & $0.159$ & $0.076$ & $0.015$ & $0.138$ & $0.059$ & $0.010$ & $0.123$ & $0.062$ & $0.012$ \\
 & $100$ & $0.125$ & $0.061$ & $0.014$ & $0.107$ & $0.052$ & $0.011$ & $0.100$ & $0.059$ & $0.016$ \\
 & $200$ & $0.115$ & $0.062$ & $0.010$ & $0.106$ & $0.059$ & $0.009$ & $0.106$ & $0.058$ & $0.010$ \\\midrule
$H_{1,1}$ & $50$ & $0.187$ & $0.093$ & $0.024$ & $0.135$ & $0.064$ & $0.010$ & $0.139$ & $0.069$ & $0.011$ \\
 & $100$ & $0.162$ & $0.094$ & $0.025$ & $0.143$ & $0.082$ & $0.025$ & $0.153$ & $0.078$ & $0.021$ \\
 & $200$ & $0.212$ & $0.121$ & $0.040$ & $0.202$ & $0.123$ & $0.033$ & $0.207$ & $0.115$ & $0.033$ \\\midrule
$H_{1,2}$ & $50$ & $0.615$ & $0.484$ & $0.180$ & $0.590$ & $0.442$ & $0.152$ & $0.551$ & $0.414$ & $0.158$ \\
 & $100$ & $0.839$ & $0.747$ & $0.509$ & $0.826$ & $0.732$ & $0.487$ & $0.810$ & $0.715$ & $0.470$ \\
 & $200$ & $0.982$ & $0.966$ & $0.892$ & $0.981$ & $0.963$ & $0.897$ & $0.980$ & $0.961$ & $0.870$ \\\midrule
$H_{1,3}$ & $50$ & $0.956$ & $0.900$ & $0.659$ & $0.948$ & $0.893$ & $0.655$ & $0.935$ & $0.873$ & $0.647$ \\
 & $100$ & $1.000$ & $0.997$ & $0.986$ & $1.000$ & $0.997$ & $0.982$ & $1.000$ & $0.996$ & $0.974$ \\
 & $200$ & $1.000$ & $1.000$ & $1.000$ & $1.000$ & $1.000$ & $1.000$ & $1.000$ & $1.000$ & $0.999$ \\\bottomrule\bottomrule
\end{tabular}
\caption{\small Empirical power of the PCvM test for the composite hypothesis $H_0: m\in \lrb{\inprod{\cdot}{\beta}:\beta\in\Hbb}$ and for different sample sizes $n$. Noise follows a $\mathcal{N}\lrp{0,0.10^2}$.\label{tab7}}
\end{table}

\subsection{Traces of the test for the simple and the composite hypothesis}

\begin{table}[H]
\centering
\small
\begin{tabular}{c|cccccc}\toprule\toprule
Models & $p=1$ & $p=2$ & $p=3$ & $p=4$ & $p=5$ & $p=6$\\\midrule
$H_{0}$ & $0.035$ & $0.037$ & $0.037$ & $0.036$ & $0.036$ & $0.037$ \\
$H_{1,2}$ & $0.044$ & $0.095$ & $0.085$ & $0.079$ & $0.076$ & $0.075$ \\
$H_{2,2}$ & $0.527$ & $0.408$ & $0.374$ & $0.362$ & $0.353$ & $0.349$ \\
$H_{3,2}$ & $0.690$ & $0.703$ & $0.705$ & $0.708$ & $0.706$ & $0.709$ \\\bottomrule\bottomrule
\end{tabular}
\caption{\small Empirical power of the PCvM test for the simple hypothesis $H_0:m(\Xcal)=\inprod{\Xcal}{\beta_0}$, $\beta_0(t)=0,\, \forall t$, for different numbers $p$ of FPC considered in the representation of the functional process. The significance level is $\alpha=0.05$ and noise follows a $\mathcal{N}\big(0,0.10^2\big)$.\label{tab3}}
\end{table}

\begin{table}[H]
\centering
\small
\begin{tabular}{c|cccccc}\toprule\toprule
Models & $p=1$ & $p=2$ & $p=3$ & $p=4$ & $p=5$ & $p=6$\\\midrule
$H_{1,0}$ & $0.063$ & $0.051$ & $0.049$ & $0.052$ & $0.061$ & $0.062$ \\
$H_{1,1}$ & $0.056$ & $0.062$ & $0.079$ & $0.078$ & $0.085$ & $0.089$ \\
$H_{1,2}$ & $0.191$ & $0.409$ & $0.686$ & $0.741$ & $0.754$ & $0.753$ \\
$H_{1,3}$ & $0.585$ & $0.908$ & $0.996$ & $0.997$ & $0.997$ & $0.997$ \\
\bottomrule\bottomrule
\end{tabular}
\caption{\small Empirical power of the PCvM test for the composite hypothesis $H_0: m\in \lrb{\inprod{\cdot}{\beta}:\beta\in\Hbb}$, for different numbers $p$ of FPC considered in the representation of the functional process. The significance level is $\alpha=0.05$ and noise is a $\mathcal{N}\big(0,0.10^2\big)$.\label{tab6}}
\end{table}

\end{document}